\documentclass[aps,prd,preprint,showpacs,superscriptaddress,showkeys,floatfix,nofootinbib]{revtex4-1}
\usepackage{graphicx}
\usepackage{dcolumn}
\usepackage{bm}
\usepackage{color}
\usepackage{url}
\usepackage{amsmath,amssymb,graphicx}
\usepackage{amsmath}
\usepackage{amssymb}
\usepackage{mathrsfs}
\usepackage{accents}
\usepackage{epsfig}
\usepackage[lofdepth,lotdepth,caption=false]{subfig}
\usepackage{mathrsfs}

\newlength\figureheight 
\newlength\figurewidth 

\begin{document}

\title{Probing light sterile neutrinos in medium baseline reactor experiments}
\author{Arman Esmaili}
\email{aesmaili@ifi.unicamp.br}
\affiliation{Instituto de F\'isica Gleb Wataghin - UNICAMP, 13083-859, Campinas, SP, Brazil}
\author{Ernesto Kemp}
\email{kemp@ifi.unicamp.br}
 \affiliation{Instituto de F\'isica Gleb Wataghin - UNICAMP, 13083-859, Campinas, SP, Brazil}
\author{O. L. G. Peres}
\email{orlando@ifi.unicamp.br}
\affiliation{Instituto de F\'isica Gleb Wataghin - UNICAMP, 13083-859, Campinas, SP, Brazil}
\affiliation{Abdus Salam International Centre for Theoretical Physics, ICTP, I-34010, Trieste, Italy}
\author{Zahra  Tabrizi}
\email{tabrizi.physics@ipm.ir}
\affiliation{School of Particles and Accelerators, Institute for Research in Fundamental Sciences (IPM), P.O.Box 19395-1795, Tehran, Iran}
\affiliation{Instituto de F\'isica Gleb Wataghin - UNICAMP, 13083-859, Campinas, SP, Brazil}

\date{\today} 

\begin{abstract}
Medium baseline reactor experiments (Double Chooz, Daya Bay and RENO) provide a unique opportunity to test the presence of light sterile neutrinos. We analyze the data of these experiments in the search of sterile neutrinos and also test the robustness of $\theta_{13}$ determination in the presence of sterile neutrinos. We show that existence of a light sterile neutrino state improves the fit to these data moderately. We also show that the measured value of $\theta_{13}$ by these experiments is reliable even in the presence of sterile neutrinos, and the reliability owes significantly to the Daya Bay and RENO data. From the combined analysis of the data of these experiments we constrain the mixing of a sterile neutrino with $\Delta m_{41}^2\sim (10^{-3}-10^{-1})~{\rm eV}^2$ to $\sin^2 2\theta_{14}\lesssim 0.1$ at $95\%$ C.L.. 
\end{abstract}

\maketitle

\section{Introduction}\label{sec:int}
Plenty of neutrino experiments performed in the last two decades confirmed that the three types of neutrinos in the Standard Model are massive particles, and the mass eigenstates $(\nu_1,\nu_2,\nu_3)$ do not coincide with the flavor eigenstates $(\nu_e,\nu_\mu,\nu_\tau)$ which enter into the charged current interactions. The mixing and flavor oscillation phenomena of neutrinos can be described by the mass-squared differences $\Delta m_{ij}^2\equiv m_i^2-m_j^2$ (where $m_i$ is the mass of $\nu_i$ state), and the so-called PMNS unitary mixing matrix parametrized by the three mixing angles $(\theta_{12},\theta_{23},\theta_{13})$ and one CP-violating phase $\delta$. The analysis of the solar and long baseline reactor neutrinos led to best-fit values $\sin^2 \theta_{12}=0.3$ and $\Delta m_{21}^2=7.4\times 10^{-5}~{\rm eV}^2$; while the data from the atmospheric and long baseline accelerator experiments result in $\sin^2\theta_{23}=0.4$ and $|\Delta m_{31}^2|\approx|\Delta m_{32}^2|=2.4\times10^{-3}~{\rm eV}^2$~\cite{GonzalezGarcia:2012sz}. The last mixing angle has been measured recently with the new generation of medium baseline reactor experiments, including Double Chooz~\cite{DCmainpaper}, Daya Bay~\cite{An:2012eh} and RENO~\cite{Ahn:2012nd}, with the best-fit value $\sin^2\theta_{13}=0.023$. 

Double Chooz, Daya Bay and RENO experiments with $L/E\sim \left( 10-10^3 \right)~{\rm m}/{\rm MeV}$, where $L$ and $E$ are the baseline and neutrino energy respectively, are sensitive to $\Delta m_{31}^2$-induced flavor oscillation with the amplitude $\sin^2 2\theta_{13}$. In fact, the measurement of the small mixing angle $\theta_{13}$ in these experiments was achieved thanks to the highly controlled systematic errors and efficient background rejection down to $\sim 10\%$ of signal. Thus, in principle, the data of these experiments can be also used to discover/constrain new physics in the neutrino sector. 

The set up and baseline to energy ratio of Double Chooz, Daya Bay and RENO experiments make them sensitive to small admixture of a new sterile neutrino state with electron anti-neutrinos, with mass-squared difference $\sim \left(10^{-3}-10^{-1}\right)~{\rm eV}^2$. The existence of a sterile neutrino state with mass $\sim 1~{\rm eV}$, the so-called $3+1$ model, was motivated by LSND~\cite{LSND}, MiniBooNE~\cite{miniboone} and reactor anomalies~\cite{reactoranomalie}. Recent global analyses of data for $3+1$ model can be found in~\cite{Kopp:2013vaa,Giunti:2013aea}. Also, several experiments have been proposed to check this scenario (see~\cite{Abazajian:2012ys} and references therein; see also~\cite{Esmaili:2013vza}). However, from the phenomenological point of view, it is worthwhile to probe the existence of a {\it light} sterile neutrino, a model which we call it $(3+1)_{\rm light}$. In this regard, we perform a detailed analysis of Double Chooz, Daya Bay and RENO data for $(3+1)_{\rm light}$ model. In this model, in addition to the $\theta_{13}$ and $\Delta m_{31}^2$ parameters, the active-sterile mixing parameters $(\theta_{14},\Delta m_{41}^2)$ contribute to $\bar{\nu}_e\to\bar{\nu}_e$ oscillation probability. We discuss the correlation among these parameters and the constraints that can be derived on them. Also, we discuss the robustness of $\theta_{13}$ determination in the presence of the sterile neutrinos. We show that the reported value of $\theta_{13}$ holds also in the presence of sterile neutrinos and the data of Daya Bay and RENO play a crucial role in this robustness. 

The prospect of sterile neutrino search in medium baseline reactor experiments has been studied in a number of papers. The possibility to search for light sterile neutrinos in reactor experiments have been mentioned in~\cite{Huber:2004xh}. In~\cite{Bandyopadhyay:2007rj}, by calculating the sensitivity of Double Chooz to sterile neutrinos with $\Delta m^2\sim 1~{\rm eV}^2$, in the $3+2$ model, it has been concluded that with only far detector data the $\theta_{13}$ angle can be confused with active-sterile neutrino mixing angles. In~\cite{deGouvea:2008qk} the interplay between a sterile neutrino with $\Delta m^2_{41} \sim (10^{-2}-1)~{\rm eV}^2$ and $\theta_{13}$ determination has been studied by computing sensitivity of Double Chooz and Daya Bay; with the conclusion that disentangling these parameters requires information about the positron recoil energy distortions. A simulation of medium baseline experiments in the search of light sterile neutrinos has been performed in~\cite{Bora:2012pi} and the dependence of limits on systematic errors has been studied. Also, in~\cite{Ciuffoli:2012yd}, the correlation of $\theta_{13}$ with active-sterile mixing parameters has been studied with an emphasize on the reactor anomaly and its connection to the cosmological data. The effect of sterile neutrinos on $\theta_{13}$ determination in both medium and long baseline experiments has been studied in~\cite{Bhattacharya:2011ee}. In~\cite{Kang:2013gpa} the data of Daya Bay and RENO has been analyzed in $3+1$ framework; the obtained limits are consistent with the limits of this paper in the same range of $\Delta m_{41}^2$. Constraining the sterile neutrino scenario with the solar and KamLAND data has been studied in~\cite{Palazzo:2012yf} which we will discuss it in section~\ref{combined}. In this paper we extend the previous searches to $\Delta m^2_{41}\sim (10^{-3}-10^{-1})~{\rm eV}^2$ and perform an analysis of the available data from the medium baseline experiments. We will show that due to the slight mismatch of data and $3\nu$ prediction at Double Chooz at $E_{\rm prompt}\sim (3-4)$~MeV, the $(3+1)_{\rm light}$ model is favored by $\sim 2.2\sigma$ significance; however, incorporating Daya Bay and RENO data diminish the significance to $\sim1.2\sigma$ C.L..

The paper is organized as follows: in section~\ref{sec:standard3nu}, we analyze the Double Chooz, Daya Bay and RENO data in the standard $3\nu$ framework. We will show that our results are consistent with theirs within the error budget. In section~\ref{sec:lightsterile}, we discuss the phenomenology of light sterile neutrinos and derive the $\bar{\nu}_e$ survival probability in $(3+1)_{\rm light}$ model. Section~\ref{sec:3p1lightsterile} is devoted to the analysis of data in $(3+1)_{\rm light}$ model. In section~\ref{analysis3p1dc} we analyze the data of Double Chooz, and in section~\ref{combined} we present the results of combined analysis of Double Chooz, Daya Bay and RENO data. Our conclusions are summarized in Section~\ref{sec:con}.

\section{Standard analysis in $3\nu$ framework}\label{sec:standard3nu}

In this section we reproduce the results of the medium baseline reactor experiments (Double Chooz~\cite{DCmainpaper}, Daya Bay~\cite{An:2012eh} and RENO~\cite{Ahn:2012nd}) in the $3\nu$ framework. Due to the moderately short baseline of these experiments, the 12-induced oscillation can be ignored and the $\bar{\nu}_e$ survival oscillation probability can be casted in the following form:     
\begin{equation}\label{psur}
P({\bar{\nu}_e\to\bar{\nu}_e})\simeq1-\sin^2 2\theta_{13}\sin^2\left( \frac{\Delta m^2_{31}L}{4E} \right),
\end{equation}
where $L\sim1~{\rm km}$ is the distance of the detector from the source, and $E\sim{\rm few~MeV}$ is the energy of the reactor neutrinos. As discussed in~\cite{DCmainpaper,An:2012eh}, in the extraction of $\theta_{13}$ value the effect of uncertainty in the value of $\Delta m_{31}^2$ is quite small and we fix it to the best-fit value $2.32\times10^{-3}~{\rm eV}^2$ measured by MINOS experiment~\cite{Adamson:2011ig}.

The detection of the reactor antineutrinos is through the Inverse Beta Decay (IBD) process, $\bar{\nu}_e +{\rm p} \to e^+ + {\rm n}$. Neutrino energy can be reconstructed by measuring the prompt positron energy $E\sim E_{{\rm prompt}}+0.78~{\rm MeV}$ (neglecting the neutron recoil energy). Thus, by reconstructing the spectrum of the observed $\bar{\nu}_e$ events, a fit of Eq.~(\ref{psur}) to the data can give information about the value of $\theta_{13}$. In the following we discuss each of these experiments in detail.   

\subsection{Double Chooz}\label{dc3analysis}

The Double Chooz experiment has detected 8,249 candidates of electron antineutrino events with 33.71 GW.ton.years exposure using a $10.3~{\rm m}^3$ detector which is located at $L=1050$~m far from the reactor cores. The total livetime of experiment is $227.93$ days. The expected number of events in the case of no-oscillation ({\it i.e.}, $\theta_{13}=0$) are 8,937 (including background events). From a rate plus spectral shape analysis they found $\sin^2 2\theta_{13}=0.109\pm 0.055$, which excludes the no-oscillation hypothesis at $99.8\%$ C.L. ($2.9~\sigma$)~\cite{DCmainpaper}. 

To reproduce the results of Double Chooz we follow the method described in~\cite{DCmainpaper}. The observed events in Double Chooz is separated in 18 prompt energy bins, between 0.7~MeV and 12.2~MeV (Fig.~14 of~\cite{DCmainpaper}). The analysis is performed by defining two different data-taking periods: {\it i}) both reactors on (139.27 days) with 6,088 total IBD candidates; {\it ii}) one reactor with less than $20\%$ of nominal power (88.66 days) with 2,161 total IBD candidates.

The numbers of expected events without oscillation, background events and observed events in each bin of energy is published in~\cite{DCmainpaper}. The expected number of events for $\theta_{13}\neq 0$ in the $i$-th bin of energy can be calculated by   
\begin{equation}\label{Nexp}
N^{{\rm osc}}_i(\theta_{13})=N^{\rm no-osc}_i\times \left<P_{{\rm sur}}(\theta_{13})\right>_i,
\end{equation}
where $N^{\rm no-osc}_i$ is the expected number of events for $\theta_{13}=0$ in the $i$-th energy bin (obtained from Fig.~14 of~\cite{DCmainpaper}, after subtracting background events), and $\left<P_{{\rm sur}}\right>_i$ is the averaged $\bar{\nu}_e$ survival probability in Eq.~(\ref{psur}) in the $i$-th energy bin. 

To analyze the data of Double Chooz we define the following $\chi^2$ function 
\begin{eqnarray}\label{chidc3}
\chi^2_{\rm DC}(\sin^2 2\theta_{13};\alpha,b)=\sum_{i=1}^{36}\frac{\left( N^{{\rm obs}}_i-\left[(1+\alpha)N^{{\rm osc}}_i(\theta_{13})+(1+b)B_i\right]\right)^2}{(\sigma^{{\rm obs}}_{i})^2+(\sigma^{{\rm osc}}_i)^{2}}
+\frac{\alpha^2}{\sigma^2_{\alpha}}+\frac{b^2}{\sigma^2_b},
\end{eqnarray}
where $i$ runs over the 36 bins of energy (18 for each period of data-taking); $N^{{\rm obs}}_i$, $B_i$ and $N^{\rm osc}_i$ are the observed, background and expected number of events in the $i$-th bin, respectively. The ${\sigma^{{\rm obs}}_{i}}=\sqrt{N^{{\rm obs}}_i}$ and ${\sigma^{{\rm osc}}_{i}}=\sqrt{N^{{\rm osc}}_i}$ represent the statistical errors of the observed events and expected events with oscillation, respectively. The systematic uncertainties in the normalization of the reactor neutrino flux and background events are taken into account by the $\alpha$ and $b$ pull terms, with $\sigma_{\alpha}=0.02$ and $\sigma_b=0.27$.
\begin{table}
\begin{tabular}{|c|c|c|c|c|}
\hline
{\rm Experiment}     & $\chi^2_{{\rm no-osc}}$ & $\chi^2_{{\rm osc}}$ &$\delta \chi^2$ &$\sin^2 2\theta_{13}$ \\
\hline
Double Chooz analysis~\cite{DCmainpaper}         & $\sim$ 52 & 42.1& 9.9 &  0.109$\pm$ 0.055 \\ 
Double Chooz (our analysis) & 35.2 & 26.2 &  9.0 & 0.115$\pm$ 0.037 \\
\hline
Daya Bay analysis~\cite{An:2012eh}                  & $\sim$ 31 & 4 & 27 & 0.092 $\pm$ 0.021\\
Daya Bay (our analysis)           &  31.8 & 3.5 & 28.3 & 0.091 $\pm$ 0.014\\
\hline
RENO analysis~\cite{Ahn:2012nd}  & $\sim$ 22 & 0 & 22 & 0.113 $\pm$ 0.032\\
RENO (our analysis) & 19.0   & 0 & 19 & 0.110  $\pm$ 0.024 \\
\hline
\end{tabular}
\caption{Comparison between our analysis and the analyses reported by Double Chooz, Daya Bay and RENO experiments. The quantity $\delta \chi^2\equiv \chi^2_{{\rm no-osc}}-\chi^2_{{\rm osc}}$ shows the improvement in the fit of data due to nonzero $\theta_{13}$.}
\label{table1}
\end{table}

\begin{figure}[t!]
\centering
\subfloat[both reactors on]{
\includegraphics[width=0.5\textwidth]{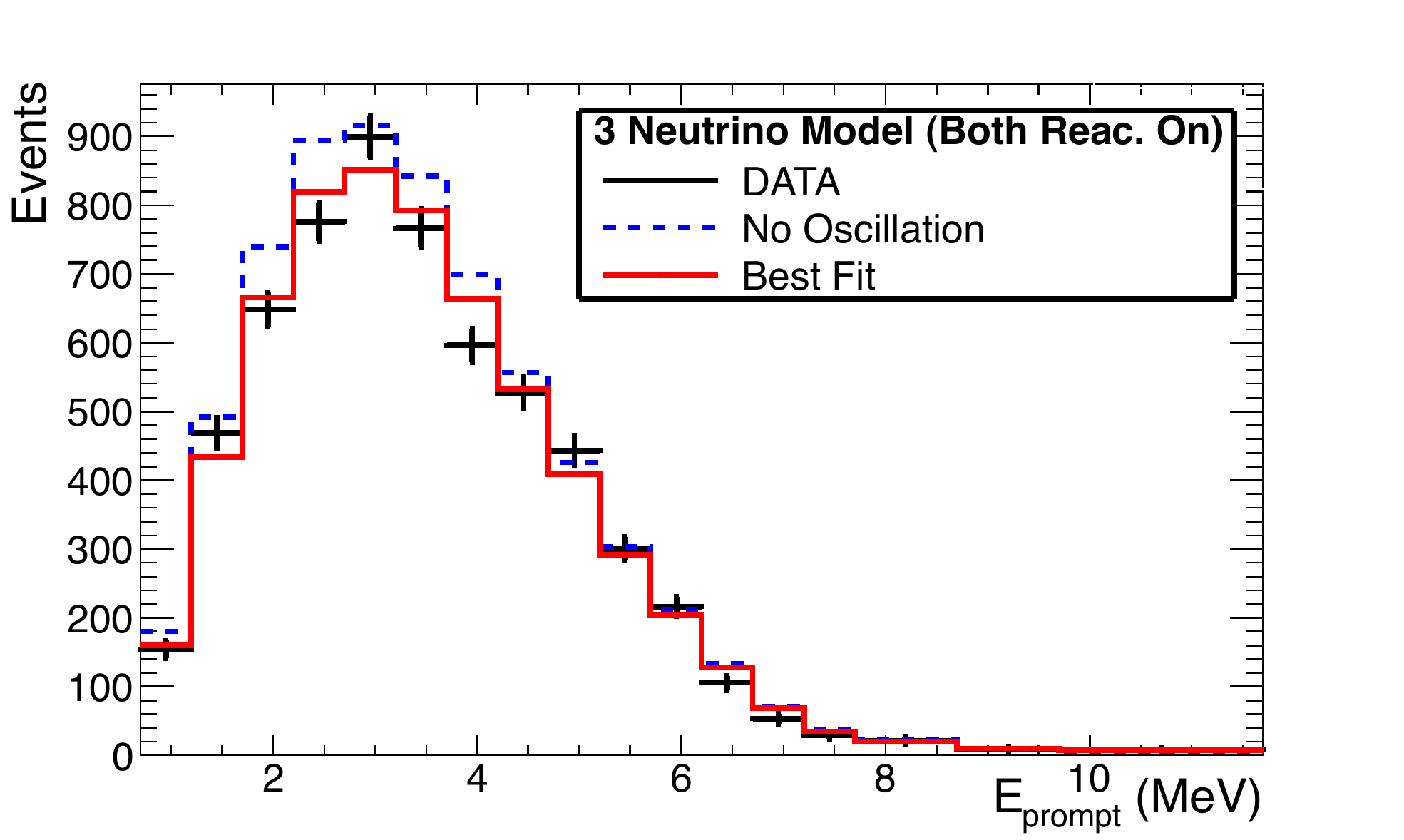}
\label{fig:dchis3,1}
}
\subfloat[one reactor off]{
\includegraphics[width=0.5\textwidth]{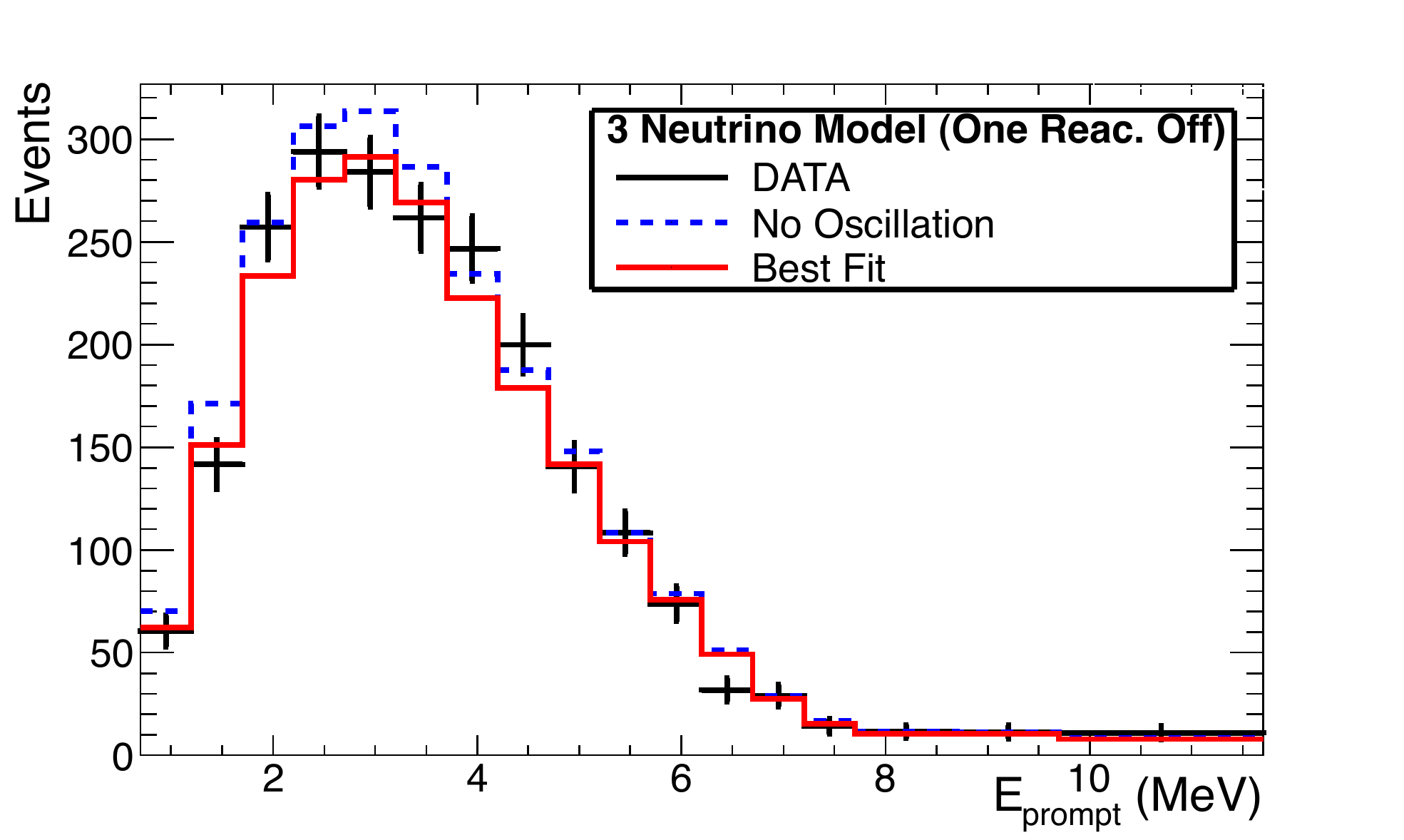}
\label{fig:dchis3,2}
}
\caption{\label{fig:dchis3}Prompt energy distribution of events in the Double Chooz experiment (data points), compared with the distributions for the best-fit value of $\theta_{13}$ (red solid curve) and $\theta_{13}=0$ (blue dashed curve). The left and right panels correspond to the two data-taking periods with both reactors on and one reactor off, respectively.}
\end{figure}

By marginalizing $\chi^2_{\rm DC}$ with respect to $\alpha$ and $b$, we obtain the best-fit value of the mixing angle $\sin^2 2\theta_{13}=0.115$, with the normalized (to the number of degrees of freedom) $\chi^2$ value of $\chi^2_{{\rm DC}}/{\rm d.o.f.}=26.2/35$, which excludes the no-oscillation hypothesis at 2.7$\sigma$ C.L. (compare to the reported 2.9$\sigma$ C.L. in~\cite{DCmainpaper}). Our $1\sigma$ range of mixing parameter $\sin^2 2\theta_{13}$ from Double Chooz is given in the first row of Table~\ref{table1}, which is similar to the range reported by Double Chooz collaboration. Also, in Fig.~\ref{fig:dchis3} we compare the best-fit energy spectrum of events from our analysis in the $3\nu$ framework (including the background events) with Double Chooz data and the spectrum for $\theta_{13}=0$, for each integration period. The contribution of the background events is $\sim15$ events in the first couple of energy bins and reduces to $\sim5$ events for higher energy bins. The left and right panels of Fig.~\ref{fig:dchis3} are for the cases where both reactors are on and one reactor runs with less than $20\%$ of power; which clearly the former plays the main role in the analysis due to the higher statistics. The preference to the nonzero $\theta_{13}$ can be easily recognized by comparing the red solid curve for the best-fit value of $\theta_{13}$ with the blue dashed curve which shows the distribution for vanishing $\theta_{13}$.

\subsection{Daya Bay}

The Daya Bay reactor neutrino experiment measured the best-fit value $\sin^2 2\theta_{13}=0.092$, excluding the zero value at 5.2$\sigma$ C.L.~\cite{An:2012eh}. In 55 days of livetime, 10,416 (80,376) electron anti-neutrino candidates have been detected in the far hall (near halls). The ratio of observed  to expected number of events is $R=0.940$. From this deficit, $\sin^2 2\theta_{13} = 0.092$ has been determined, based on a rate-only analysis.

The Daya Bay experiment consists of three underground experimental halls (EH1, EH2 and EH3), where two Antineutrino Detectors (AD) are located in EH1 and one AD in EH2 (near halls). Three ADs are located at the far hall (EH3) at a distance where $\bar{\nu}_e$ survival oscillation probability deviates maximally from one. Daya Bay collaboration have published the ratio of observed to expected number of events (assuming no-oscillation) for each AD (Fig.~4 of~\cite{An:2012eh}); {\it i.e.}, 
\begin{equation}\label{ratio}
R_i\equiv\frac{\#{\rm ~of~observed~events~in~}i{\rm -th~AD}}{\#{\rm ~of~expected~events~in~}i{\rm -th~AD}}~,
\end{equation}
where $i=1,\ldots, 6$. Using $R_i$s and the number of observed IBD candidates in each AD (given in Table II of~\cite{An:2012eh}), the expected number of events without oscillation can be calculated by: $N^{\rm no-osc}_i=N_i^{\rm obs}/R_i$. The averaged oscillation probability for each AD of Daya Bay can be written as (see~\cite{Ciuffoli:2012yd})
\begin{equation}
\langle P_i\rangle=1-\sin^2 2\theta_{13} \int \sin^2 \left(\frac{\Delta m^2_{31}d_i}{4E}\right)\rho(E) {\rm d}E~,
\label{eqpi}
\end{equation} 
where $d_i$ is the weighted distance of the $i$-th AD from the reactors (Fig.~4 of~\cite{An:2012eh}) and $\rho(E)$ is the fractional energy distribution of neutrinos. Using the top panel of Fig.~5 in~\cite{An:2012eh} for the energy distribution of events, the expected number of events including the oscillation for the $i$-th AD becomes
\begin{equation}\label{expected}
N^{\rm exp}_i=\left(\dfrac{N_i^{{\rm obs}}}{R_i}\right)\times \langle P_i\rangle~.
\end{equation}

We analyze the Daya Bay data by defining the following $\chi^2$ function (which includes only the rate information):
 \begin{eqnarray}
 \chi^2_{{\rm DB}}(\sin^2 2\theta_{13};\epsilon,\epsilon_i,\alpha_r,\eta_i)&=&\sum_{i=1}^6 \frac{\left[N_i^{{\rm obs}}-N^{{\rm exp}}_i(\theta_{13})\left\{1+\epsilon+\sum_r \omega_r^i \alpha_r+\epsilon_i\right\}+\eta_i\right]^2}{
\left(\sigma_i^{{\rm stat}}\right)^2} \nonumber \\
  &+&\sum_r\frac{\alpha_r^2}{\sigma_r^2}+\sum_{i=1}^6\left(\frac{\epsilon^2_i}{\sigma^2_d}+\frac{\eta^2_i}{(\sigma^2_B)_i}\right)~,
 \label{chidn}
 \end{eqnarray}
where $N_i^{{\rm obs}}$ and $N_i^{{\rm exp}}$ are respectively the total number of observed and expected IBD candidate events in the $i$-th AD; with $\sigma_i^{{\rm stat}}$ representing statistical error of observed number of events which is defined as $(\sigma_i^{{\rm stat}})^2=N_i^{{\rm obs}}+B_i$, where $B_i$ is the number of background events in the $i$-th AD. The $\omega^i_r$ is the fraction of neutrino flux from $r$-th reactor at the $i$-th AD. The systematic uncertainties of reactor flux, detection efficiency and background events are taken into account by the pull terms with the pull parameters $\alpha_r$, $\epsilon_i$ and $\eta_i$ respectively; with $\sigma_r=0.8\%$, $\sigma_d=0.2\%$ and $(\sigma_B)_i$ presented in Table II of~\cite{An:2012eh}. The parameter $\epsilon$ accommodates the uncorrelated flux normalization uncertainty which we marginalize without any pull term compensation. All the values used in $\chi^2_{\rm DB}$ function of Eq.~(\ref{chidn}) are listed in Table II of~\cite{An:2012eh}.

After minimizing $\chi^2_{\rm DB}$ with respect to all parameters, we find the best-fit value of the mixing parameter $\sin^2 2\theta_{13}=0.091$, which is consistent with the Daya Bay result. The $\chi^2$ value and $1\sigma$ range of $\sin^2 2\theta_{13}$ are shown in the second row of Table~\ref{table1}.

\subsection{RENO}

The RENO experiment observed disappearance of reactor $\bar{\nu}_e$ with $4.9\sigma$ of significance. In 229 days of data-taking period, the number of observed neutrinos at far (near) detector is 17,102 (154,088). The ratio of the number of observed neutrinos to the number of expected neutrinos (for $\theta_{13}=0$) is $R=0.920$. From this deficit, RENO collaboration obtained $\sin^2 2\theta_{13} = 0.113$, based on a rate-only analysis~\cite{Ahn:2012nd}.

The RENO experiment consists of two detectors, near and far, detecting $\bar{\nu}_e$ emission form   6 reactors. The average distance of the near (far) detector from the center of reactor array is 294 m (1383 m). The details of the experiment can be found at~\cite{Ahn:2012nd}. The analysis of RENO is similar to the analysis of Daya Bay. Following the method described in~\cite{Ciuffoli:2012yd}, we calculate the averaged survival probability of neutrinos in $i$-th detector ($i=$~near,far) in the following way:
\begin{eqnarray}
\langle P_i\rangle=\sum_{j=1}^6f_{ij}\left[1-\sin^2 2\theta_{13}\int \sin^2\left(\frac{\Delta m^2_{31}d_{ij}}{4E}\right)\rho_j(E) {\rm d}E\right],~~~~~
\label{psurvRENO}
\end{eqnarray}
where $f_{ij}$ is the fraction of antineutrino flux at the $i$-th detector coming from the $j$-th reactor, given in Table~1 of~\cite{Ahn:2012nd}; $d_{ij}$ is the weighted distance of the $i$-th detector from the $j$-th reactor and $\rho_j$ is the fractional energy distribution of neutrinos emitted from the $j$-th reactor. Using the ratio of observed to expected number of events (as defined in Eq.~(\ref{ratio})) from the bottom panel of figure~3 in~\cite{Ahn:2012nd}, the expected number of events without oscillation can be derived. The expected number of events including oscillation can be calculated form Eq.~(\ref{expected}).

We use the same $\chi^2$ function defined by the collaboration~\cite{Ahn:2012nd}:
\begin{eqnarray}\label{chiRENO}
\chi^2_{{\rm RENO}}(\sin ^2 2\theta_{13};\alpha,b_i,\xi_i,f_r)=\sum_{i=N,F} \frac{\left[N_i^{{\rm obs}}+b_i-(1+\alpha+\xi_i)\sum_{r=1}^6(1+f_r)N_{i,r}^{\rm exp}\right]^2}{\left(\sigma_i^{{\rm obs}}\right)^2}\nonumber\\
+\sum_{i=N,F}\left(\frac{\xi^2_i}{\sigma_d^{\xi^2}}+\frac{b^2_i}{\sigma^{b^2}_i}\right)+\sum_{r=1}^6\left(\frac{f_r}{\sigma_r}\right)^2,~~~~~~~~~~~~~~~~~~~~~~~~~
\end{eqnarray}
where $i$ denotes either Near or Far detectors, $N_i^{{\rm obs}}$ is the total number of neutrinos observed in each detector (after background subtraction), $r=1,\ldots,6$ runs over the reactors and $\alpha$ takes into account the global flux normalization uncertainty. $\sigma_i^{{\rm obs}}$ is the statistical error of observed events; and the uncorrelated systematic error of reactor flux and detection efficiency are $\sigma_r=0.9\%$ and $\sigma _d^\xi=0.2\%$ respectively (taken from Table II of~\cite{Ahn:2012nd}). $\sigma^b_{N(F)}$ are the background uncertainties at near (far) detectors (listed in Table I of~\cite{Ahn:2012nd}). The corresponding pull parameters are respectively $f_r$, $\xi_i$ and $b_i$.

\begin{figure}[t!]
\centering
\includegraphics[width=0.6\textwidth]{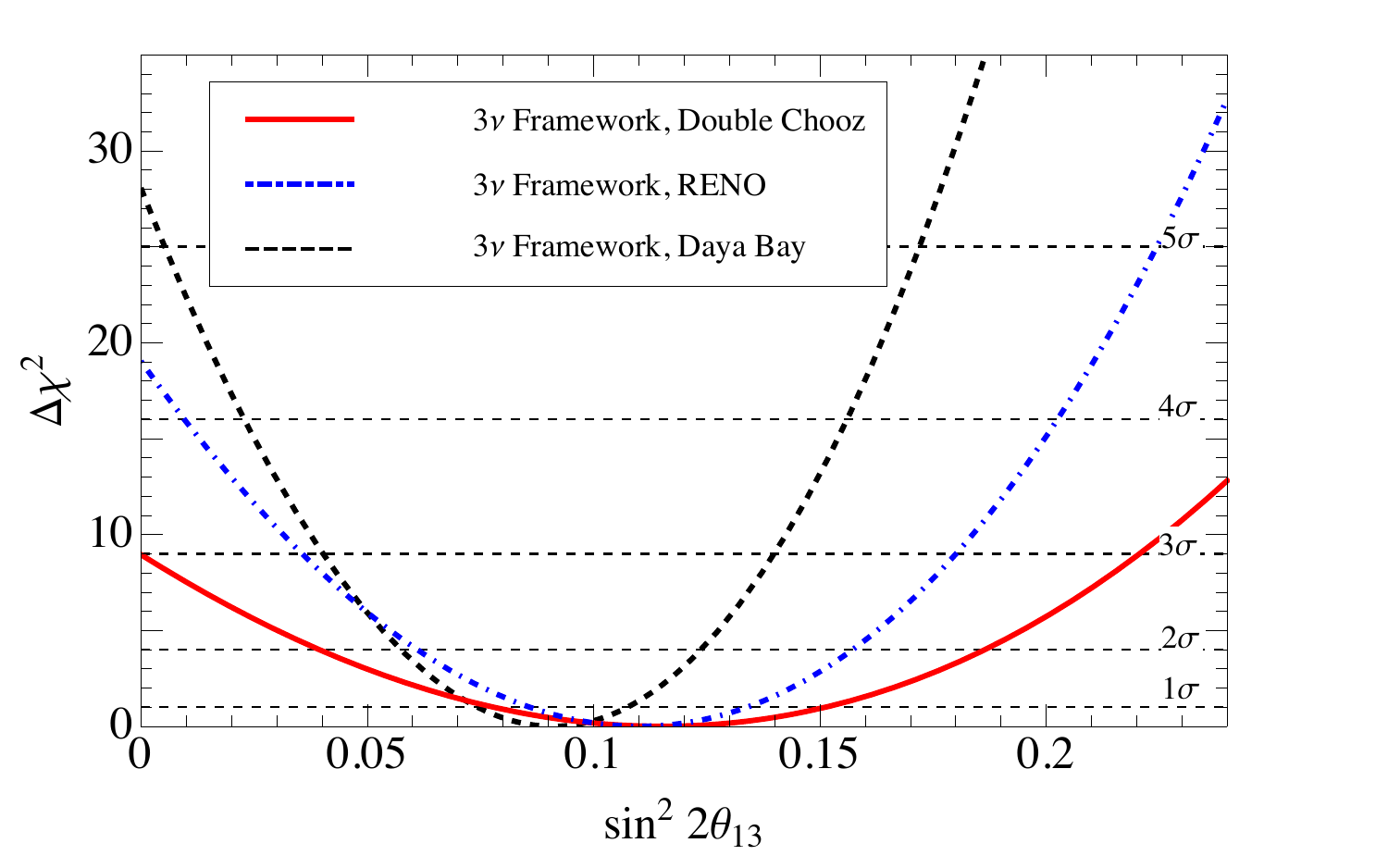}
\caption{\label{fig:chi2,all,3nu}$\Delta \chi^2\equiv \chi^2 - \chi^2_{\rm min}$ versus $\sin^2 2\theta_{13}$, for Double Chooz (red solid curve), RENO (blue dot-dashed curve) and Daya Bay (black dashed curve) experiments.}
\end{figure}

Minimizing $\chi^2_{{\rm RENO}}$ of Eq.~(\ref{chiRENO}) with respect to all the pull parameters, we find the best-fit value $\sin^2 2\theta_{13}=0.110$, which is consistent with the RENO result reported in~\cite{Ahn:2012nd}. The summary of all our results about RENO can be seen in the third row of Table~\ref{table1}.

In Fig.~\ref{fig:chi2,all,3nu} we show $\Delta\chi^2\equiv \chi^2-\chi^2_{\rm min}$ versus $\sin^2 2\theta_{13}$ for the three discussed experiments. Comparing our results with the results of the Double Chooz, Daya Bay and RENO collaborations in Table~\ref{table1} show that: $i$) our best-fit values for $\sin^2 2\theta_{13}$ are fairly close to the reported values by collaborations; $ii$) our exclusion of $\theta_{13}=0$ (quantified by $\delta \chi^2$) and also $1\sigma$ allowed interval of $\sin^2 2\theta_{13}$ are compatible with the corresponding values reported by the collaborations. It worths to mention that the errors in the value of $\sin^22 \theta_{13}$ obtained in our analyses are moderately smaller than the errors obtained by the collaborations (see Table~\ref{table1}). This can be due to some optimistic assumptions and/or lack of information about the systematics. However, the closeness of our best-fit values and fair matching of error intervals to the collaboration values, make our analysis viable. In the rest of this paper, we will use the data of these experiments to probe the existence of sterile neutrino state with mass-squared difference $\Delta m^2\sim (10^{-3}-10^{-1})\,{\rm eV}^2$.

\section{Framework of light sterile neutrino: $(3+1)_{\rm light}$}\label{sec:lightsterile}

Although the majority of data from the neutrino oscillation experiments can be interpreted consistently in the $3\nu$ framework, persisting anomalies (including MiniBooNE, LSND, reactor and Gallium anomalies) motivate the existence of a sterile neutrino state with mass $\sim \mathcal{O}(1)$~eV. Although several experiments have been proposed to check the existence of $\sim 1$~eV sterile neutrinos (see~\cite{Abazajian:2012ys,Esmaili:2013vza}), the possibility of the existence of a light sterile neutrino ($\Delta m^2_{41}\ll 1~$eV$^2$) is neither excluded strongly nor planned to be explored substantially\footnote{The potential of upcoming reactor experiments with larger baseline to probe ``super light" sterile neutrinos ($\Delta m_{41}^2\simeq10^{-5}~{\rm eV}^2$) is studied in \protect{\cite{Bakhti:2013ora}}.}. In this paper we probe this possibility in the light of Double Chooz, Daya Bay and RENO published data.

For simplicity, we extend the neutrino sector of the Standard Model by adding one light sterile neutrino: the $(3+1)_{\rm light}$ model. In this model the mass spectrum of neutrino sector consists of three mostly active neutrino mass eigenstates with masses $(m_1,m_2,m_3)$ and one mostly sterile neutrino mass eigenstate with mass $m_4$ such that: 
\begin{equation}
m_1<m_2 \ll m_3<m_4~.
\end{equation}
The mixing of these states can be described by generalizing the PMNS matrix to a $4\times4$ unitary matrix, $U_4$, which can be parametrized by six mixing angles (we assume CP-conservation in neutrino sector). Specifically, we adopt the following parametrization of $U_4$:
\begin{eqnarray}
U_{4}= R^{34}(\theta_{34})R^{24}(\theta_{24})R^{14}(\theta_{14})R^{23}(\theta_{23})R^{13}(\theta_{13})R^{12}(\theta_{12}) ,
\label{uei}
\end{eqnarray}
where $R^{ij}(\theta_{ij})$ (i,j = 1,...,4 and $i < j$) is the $4\times4$ rotation matrix in the $ij$-plane with angle $\theta_{ij}$. Totally, four new mixing parameters are introduced in $(3+1)_{\rm light}$ model: three mixing angles $(\theta_{14},\theta_{24},\theta_{34})$ which quantify the $\nu_s-\nu_e$, $\nu_s-\nu_\mu$ and $\nu_s-\nu_\tau$ mixings, respectively; and one new mass-squared difference which we choose as $\Delta m_{41}^2\equiv m_4^2-m_1^2$. The two extra mass-squared differences $\Delta m_{42}^2$ and $\Delta m_{43}^2$ are not independent and can be written as: $\Delta m^2_{42}=\Delta m^2_{41}-\Delta m^2_{21}$ and $\Delta m^2_{43}=\Delta m^2_{41} -\Delta m^2_{31}$.

\begin{figure}[t!]
\centering
\subfloat{
\includegraphics[width=0.5\textwidth]{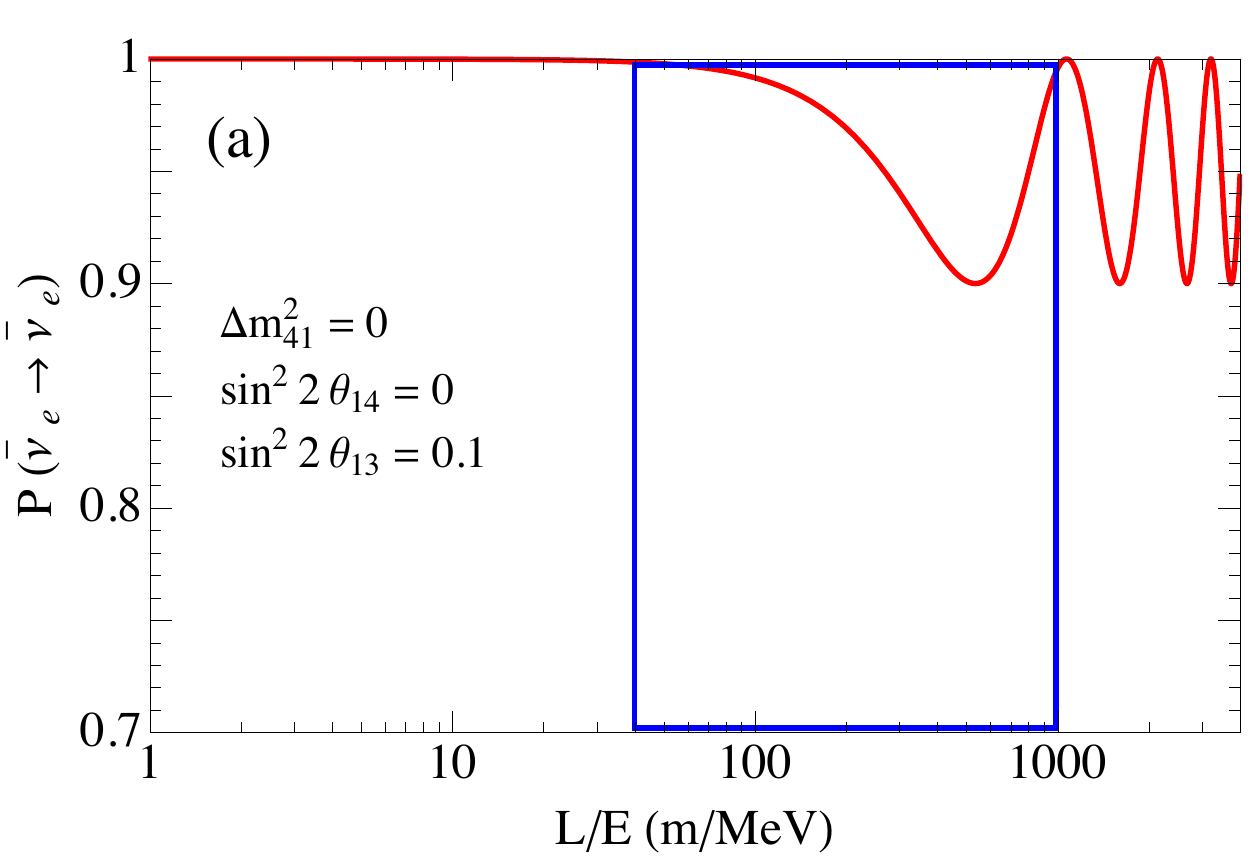}
\label{fig:prob11}
}
\subfloat{
\includegraphics[width=0.5\textwidth]{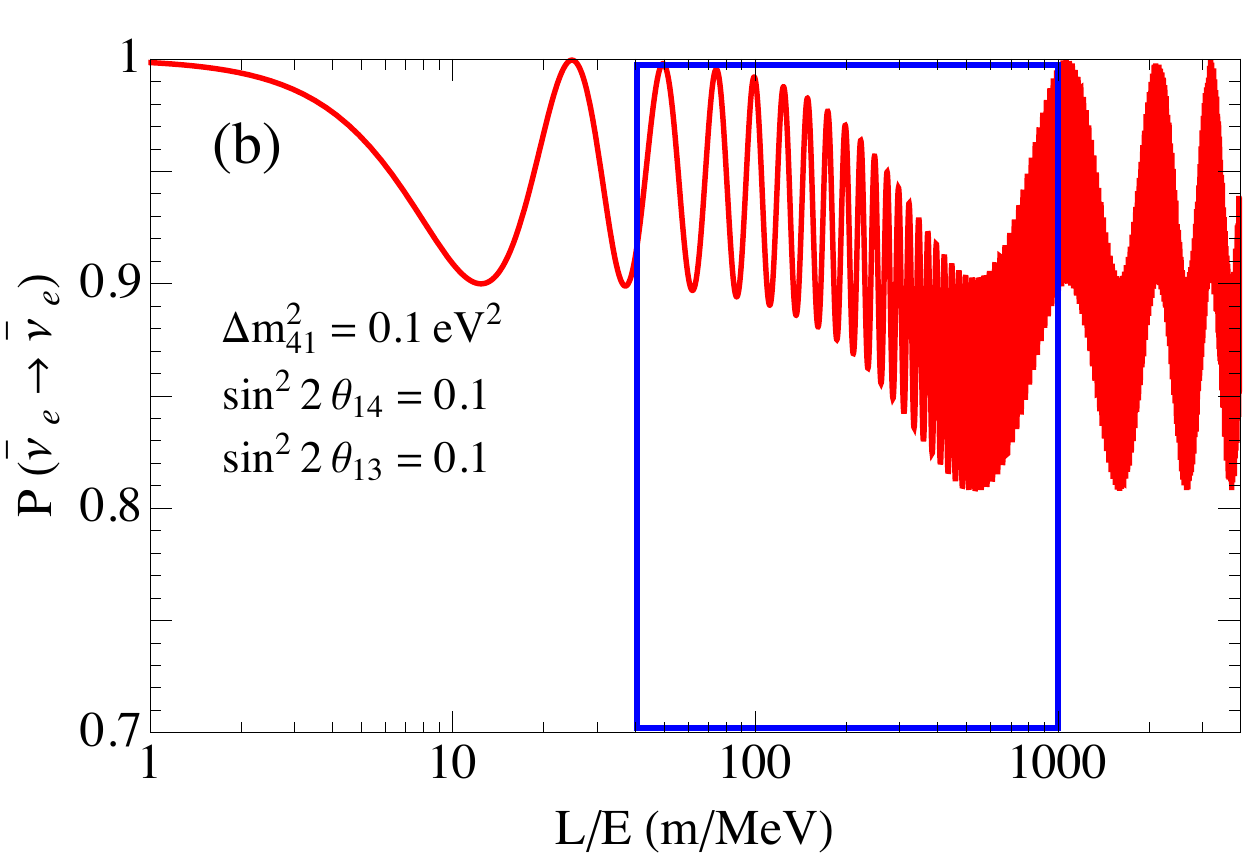}
\label{fig:prob2}
}
\quad
\subfloat{
\includegraphics[width=0.5\textwidth]{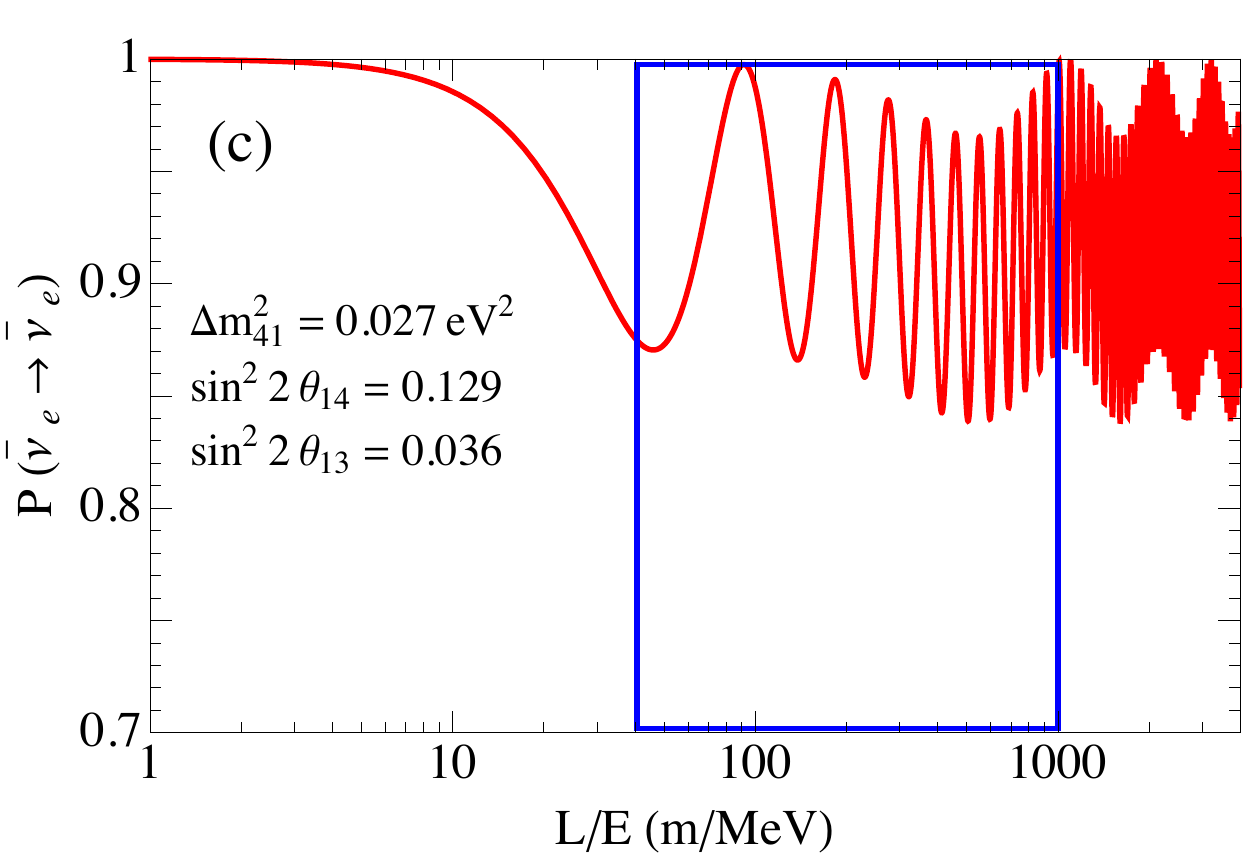}
\label{fig:prob3}
}
\subfloat{
\includegraphics[width=0.5\textwidth]{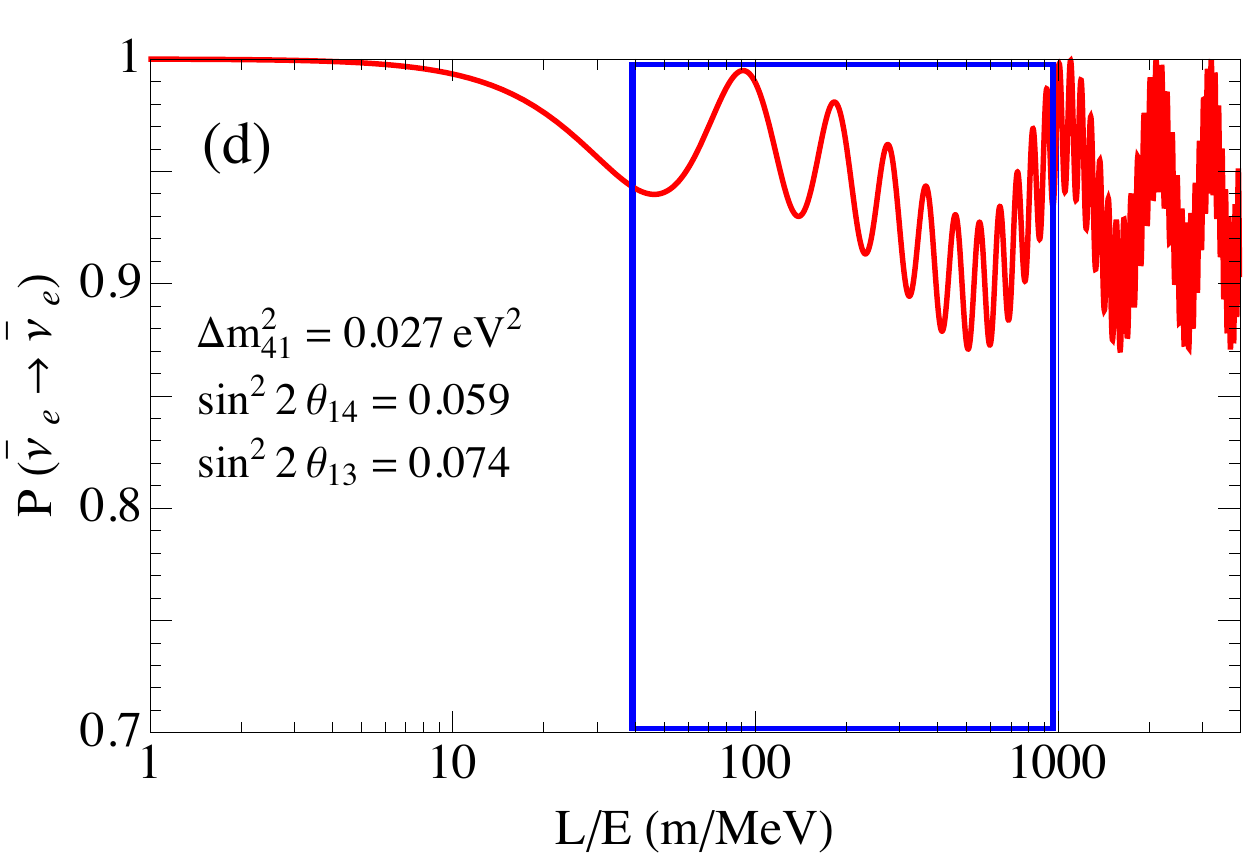}
\label{fig:prob4}
}
\caption{\label{fig:prob,3+1}The $\bar{\nu}_e$ survival probability versus $L/E$ for different values of $\sin^2 2\theta_{13}$, $\sin^2 2\theta_{14}$ and $\Delta m_{41}^2$. The blue box represents the relevant $L/E$ region for the medium baseline experiments. Panel (a) shows the probability in the $3\nu$ framework with the best-fit value of $\theta_{13}$. Panel (b) is for nonzero active-sterile mixing parameters, with a rather large $\Delta m_{41}^2$. Panels (c) and (d) show the probability for the best-fit values obtained in the analysis of $(3+1)_{\rm light}$ model, for Double Chooz and the combined data, respectively.}
\end{figure}

In the medium baseline reactor experiments (Double Chooz, Daya Bay and RENO) which we are considering in this paper, the distances of near and far detectors to the sources are a few hundred meters and $\sim1$~km, respectively; and the energy of neutrinos emitted from reactors is $\sim$ a few MeV. In this energy and baseline ranges the oscillation phase induced by $\Delta m_{21}^2$ can be ignored and the $\bar{\nu}_e$ survival probability is given by:
\begin{eqnarray}
P({\bar{\nu}_e\to\bar{\nu}_e})&=&1-4\left(|U_{e1}^2|+|U_{e2}|^2\right)\times\left[|U_{e3}|^2\sin^2
\left(\frac{\Delta m^2_{31}L}{4E}\right) \right.\nonumber\\
 & + &\left. |U_{e4}^2|\sin^2\left(\frac{\Delta m^2_{41}L}{4E}\right)\right] -4|U_{e3}|^2|U_{e4}|^2\sin^2\left(\frac{\Delta m^2_{43}L}{4E}\right)~,
\end{eqnarray}
where $L$ and $E$ are the baseline and energy respectively. In terms of the parametrization of Eq.~(\ref{uei}), the oscillation probability is
\begin{eqnarray}\label{3+1probability}
P(\bar{\nu}_e\to\bar{\nu}_e)&=&1-\sin^22\theta_{13} \cos^4\theta_{14}\sin^2\left(\frac{\Delta m^2_{31}L}{4E}\right) - \cos^2\theta_{13}\sin^2 2\theta_{14}\sin^2\left(\frac{\Delta m^2_{41}L}{4E}\right) \nonumber \\
 & - & \sin^2\theta_{13}\sin^2 2\theta_{14}\sin^2\left(\frac{\Delta m^2_{43}L}{4E}\right)~.
\end{eqnarray}
In the analysis of $(3+1)_{\rm light}$ scenario, we treat the $\theta_{13}$ , $\theta_{14}$ and $\Delta m_{41}^2$ in Eq.~(\ref{3+1probability}) as free parameters and fix $\Delta m_{31}^2$ to its best-fit value from MINOS experiment~\cite{Adamson:2011ig}. In Fig.~\ref{fig:prob,3+1} we show the $\bar{\nu}_e$ survival probability for different values of $\sin^2 2\theta_{14}$, $\sin^2 2\theta_{13}$ and $\Delta m_{41}^2$, versus $L/E$. The blue box shows the relevant values of $L/E$ for the medium baseline experiments for both near and far detectors. The upper left panel shows the probability for the best-fit value of $\theta_{13}$ in the $3\nu$ framework. The upper right panel depicts the probability for nonzero active-sterile mixing parameters shown in the legend; and as can be seen, for relatively large $\Delta m_{41}^2=0.1~{\rm eV}^2$ in this panel the effect of sterile neutrino is averaged out especially for the far detector. The two lower panels show the probability for the best-fit values of the mixing parameters obtained in our analysis of $(3+1)_{\rm light}$ (see section~\ref{sec:3p1lightsterile}).

Some symmetries in Eq.~(\ref{3+1probability}) can be recognized. For example, there is a degeneracy in Eq.~(\ref{3+1probability}) where in the limit $\Delta m_{43}^2\to0$ (or $\Delta m_{41}^2\to\Delta m_{31}^2$), when $\theta_{13}=0$, the angle $\theta_{14}$ imitates the role of $\theta_{13}$ when $\theta_{14}=0$. Thus, it is always possible to obtain a fit to the data in $(3+1)_{\rm light}$ model as good as the fit in $3\nu$ framework by setting $\Delta m_{41}^2=\Delta m_{31}^2$ and exchanging $\theta_{14}$ with $\theta_{13}$. However, it would be possible that deviation of $\Delta m_{41}^2$ from $\Delta m_{31}^2$ (that is, nonzero $\Delta m_{43}^2$), which leads to a shift of extrema positions in Fig.~\ref{fig:prob,3+1}, results in a better fit than the $3\nu$ fit. Notice that varying $\theta_{13}$ in the $3\nu$ scheme leads to a change in the depth of minima while leaving the positions of minima unchanged. So, generally we expect to obtain better fits by extending the $3\nu$ framework to $(3+1)_{\rm light}$ model. The possibility of having two independent measurements in near and far detectors (as in Daya Bay and RENO) helps in lifting this degeneracy (see section~\ref{combined}), while with one detector only (Double Chooz) the degeneracy manifests (see section~\ref{analysis3p1dc}).

\section{Analysis for $(3+1)_{\rm light}$ model}\label{sec:3p1lightsterile}
In this section we confront the data from Double Chooz, Daya Bay and RENO experiments with the prediction of $(3+1)_{\rm light}$ model. In section~\ref{analysis3p1dc}, we consider the Double Chooz data and in section~\ref{combined}, we perform the combined analysis including the data of the other two experiments.   

\subsection{Probing $(3+1)_{\rm light}$ model with Double Chooz}
\label{analysis3p1dc}

For the analysis of the Double Chooz data for $(3+1)_{\rm light}$ model, we follow the method of section~\ref{dc3analysis}, with the modification that the number of expected events in the $(3+1)_{\rm light}$ model is given by 
\begin{equation}
N^{{\rm osc,3+1}}_i=N^{{\rm no-osc}}_i \times \left\langle P(\theta_{13},\theta_{14},\Delta m^2_{41} )\right\rangle_i~,
\end{equation}
where $\left\langle P\right\rangle_i$ is the average of survival probability in Eq.~(\ref{3+1probability}) in the $i$-th energy bin.

Using the same $\chi^2_{\rm DC}$ as in Eq.~(\ref{chidc3}), we find the following best-fit values: 
\begin{equation}
\sin^2 2\theta_{13}=0.036\quad , \quad\sin^2 2\theta_{14}=0.129\quad , \quad\Delta m^2_{41}=0.027~{\rm eV}^2~,
\label{3p1bf}
\end{equation}
with the minimum value $\chi^2_{{\rm DC}}/{\rm d.o.f}=19.1/33$, which shows improvement with respect to $3\nu$ analysis with minimum of $\chi^2_{{\rm DC}}/{\rm d.o.f}=26.2/35$ (see the first row of Table~\ref{table1}). The main feature of the $(3+1)_{\rm light}$ analysis with Double Chooz data is the significantly different best-fit value for $\sin^2 2\theta_{13}$ in Eq.~(\ref{3p1bf}) with respect to the $3\nu$ best-fit value $\sin^2 2\theta_{13}=0.115$. In Fig.~\ref{fig:chi2,dc3+1,q13}, we compare the $\chi^2_{\rm DC}$ as a function of $\sin^2 2\theta_{13}$ for the $3\nu$ and $(3+1)_{\rm light}$ models. The red dashed curve for $(3+1)_{\rm light}$ is calculated by marginalizing $\chi^2_{\rm DC}$ with respect to $\Delta m_{41}^2$ and $\sin^2 2\theta_{14}$. The shift in the best-fit value of $\sin^2 2\theta_{13}$ in $(3+1)_{\rm light}$ is clear from Fig.~\ref{fig:chi2,dc3+1,q13}. Also, from the best-fit value $\sin^22\theta_{13}=0.036$ down to zero, the $\chi^2$ value is nearly constant which shows the negative impact of $\theta_{14}$ on establishing a nonzero value for $\theta_{13}$ from Double Chooz data. Thus, as we expect, for the Double Chooz data active-sterile mixing parameters can mimic the effect of $\theta_{13}$ (see section~\ref{sec:lightsterile}).

\begin{figure}[t!]
\centering
\includegraphics[width=0.55\textwidth]{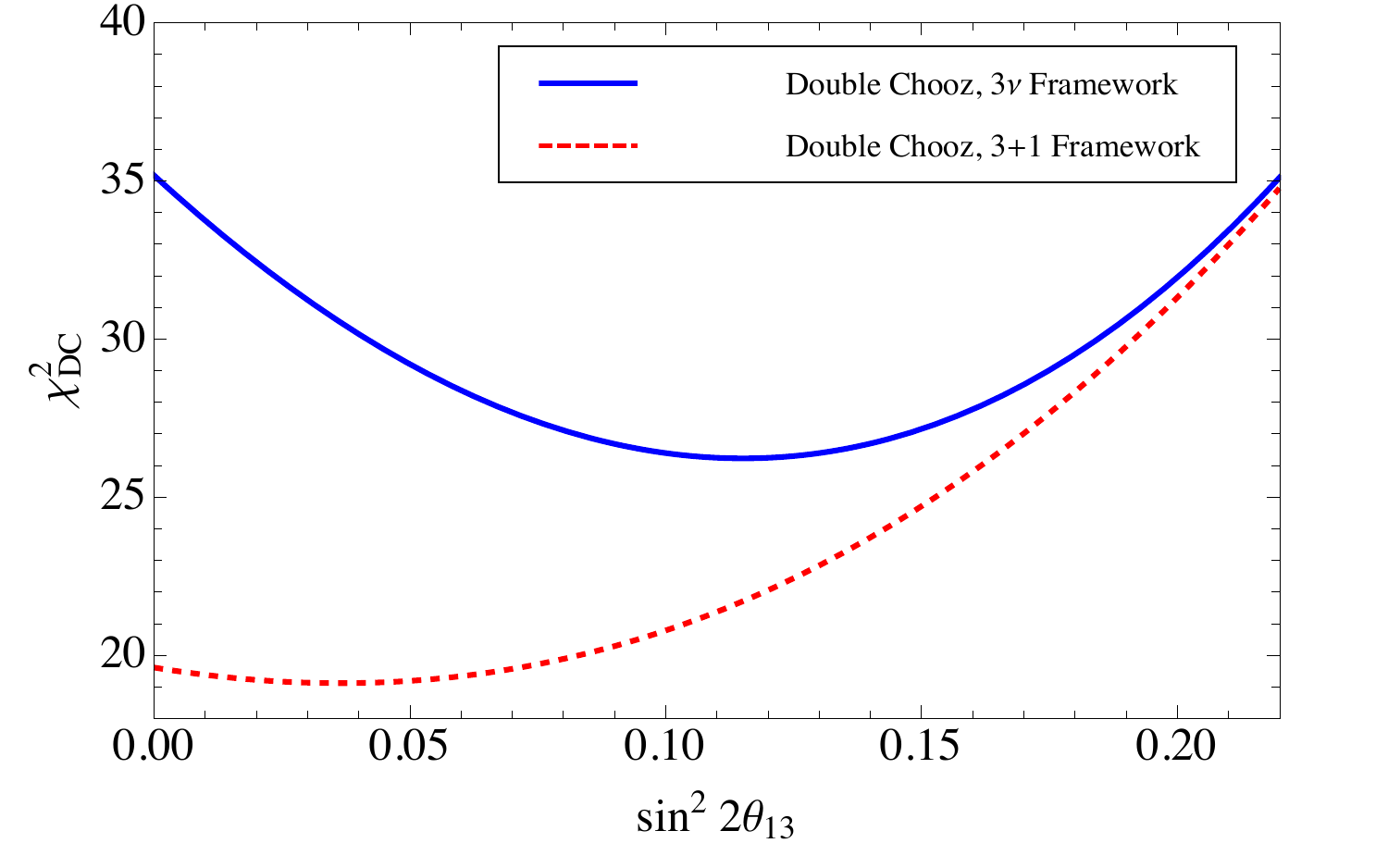}
\label{fig:prob1}
\caption{\label{fig:chi2,dc3+1,q13}$\chi^2_{{\rm DC}}$ versus $\sin^2 2\theta_{13}$ for $(3+1)_{\rm light}$ model (red dashed curve) and $3\nu$ framework (blue solid curve), for Double Chooz data.}
\end{figure}

\begin{figure}[t!]
\centering
\subfloat[]{
\includegraphics[width=0.5\textwidth]{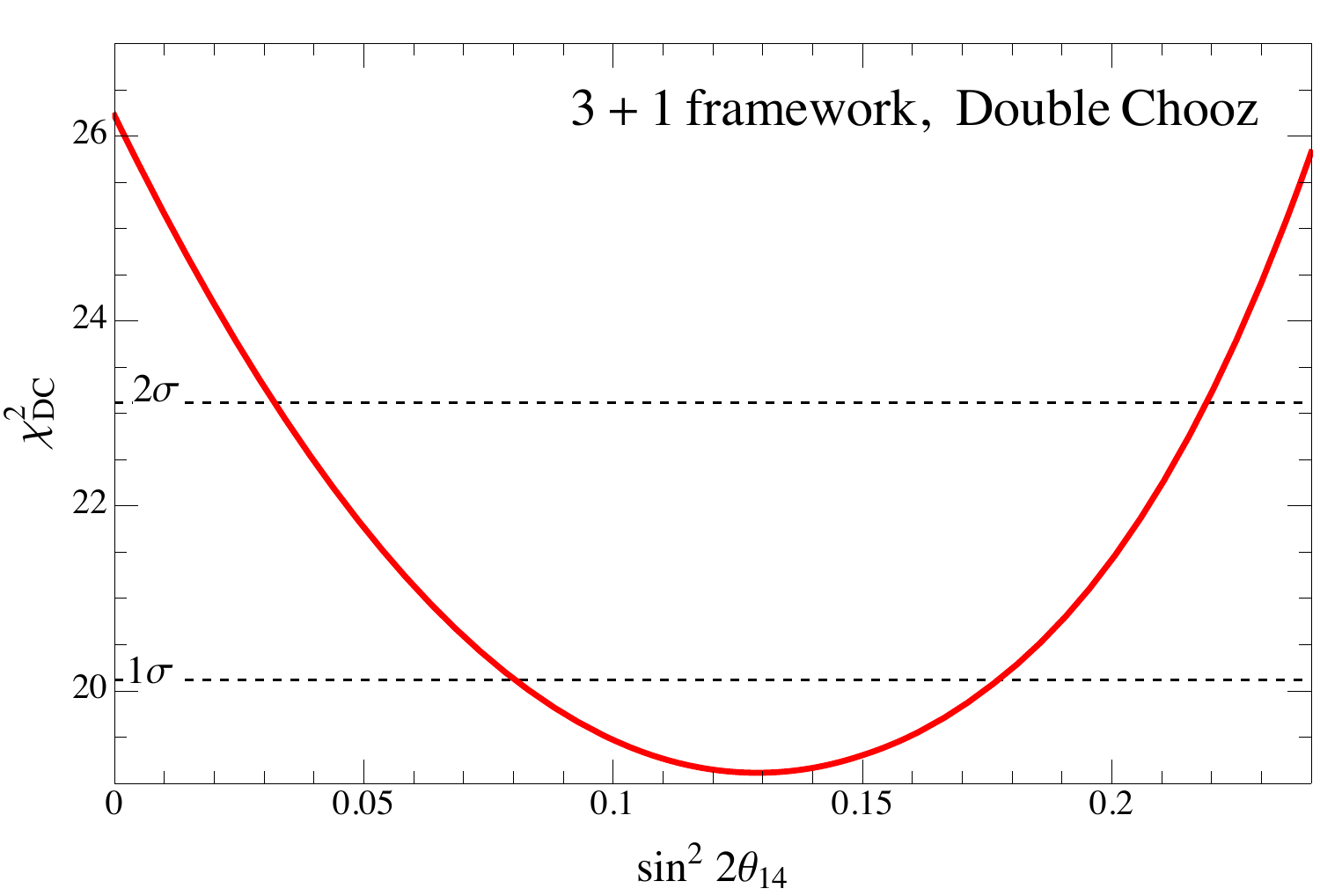}
\label{fig:chi2,dc3+1,1}
}
\subfloat[]{
\includegraphics[width=0.5\textwidth]{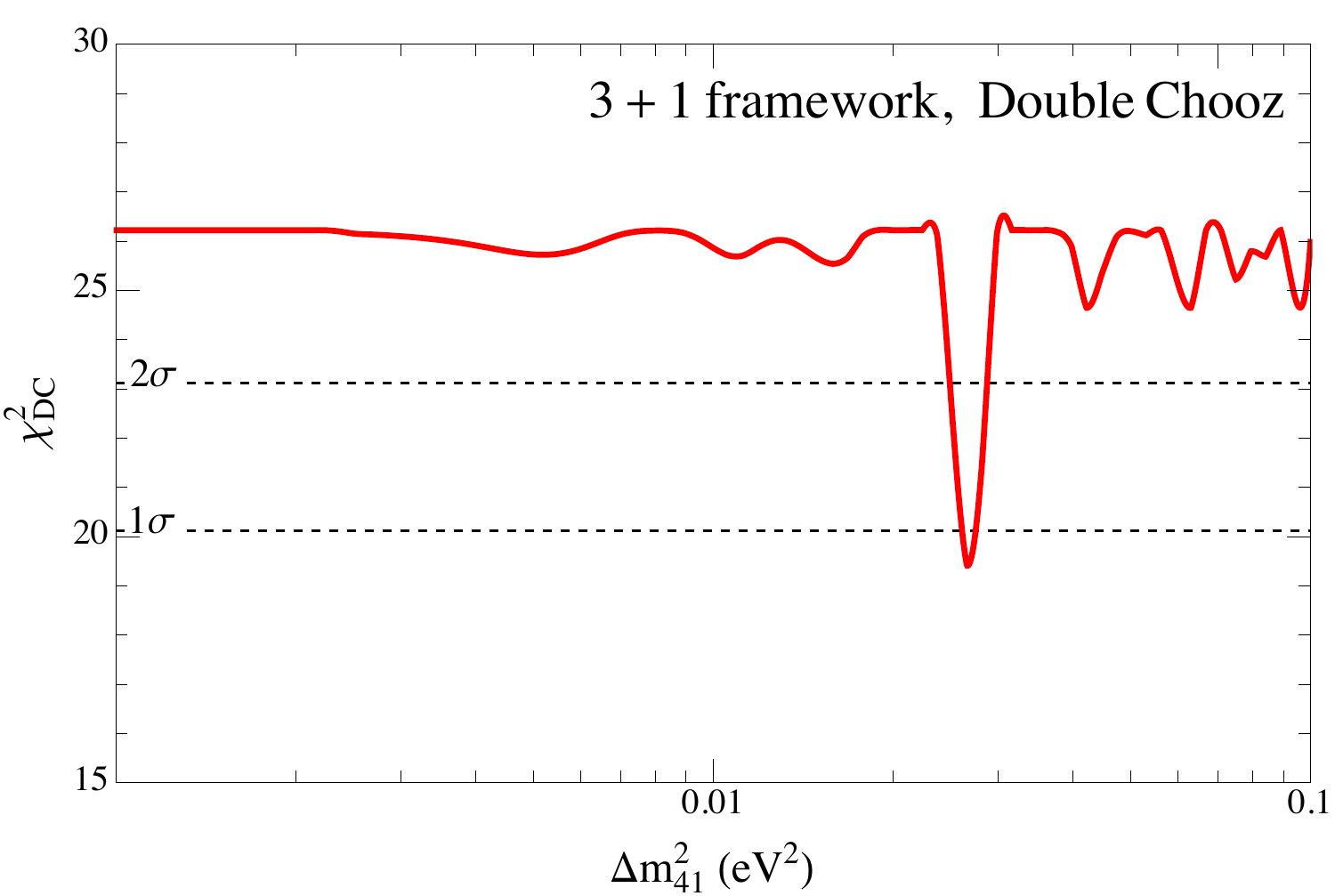}
\label{fig:chi2,dc3+1,2}
}
\caption{\label{fig:chi2,dc3+1}$\chi^2_{{\rm DC}}$ with respect to $\sin^2 2\theta_{14}$ (left panel) and $\Delta m_{41}^2$ (right panel), which are calculated by marginalizing $\chi^2_{\rm DC}$ with respect to $\Delta m_{41}^2$ and $\sin^2 2\theta_{13}$ for the left panel, and $\sin^2 2\theta_{13}$ and $\sin^2 2\theta_{14}$ for the right panel.}
\end{figure}

In the left and right panels of Fig.~\ref{fig:chi2,dc3+1} we show the $\chi^2_{\rm DC}$ values as a function of $\sin^2 2\theta_{14}$ and $\Delta m_{41}^2$, respectively; where the $\chi^2_{\rm DC}$ values are obtained by marginalization over other parameters. As can be seen from Fig.~\ref{fig:chi2,dc3+1,1}, with the Double Chooz data, zero value of $\sin^2 2\theta_{14}$ can be excluded by more than $\sim $ 2.2$\sigma$. Comparing Fig.~\ref{fig:chi2,dc3+1,1} with the red curve in Fig.~\ref{fig:chi2,all,3nu} shows that, as we anticipated, for Double Chooz data extending $3\nu$ framework to $(3+1)_{\rm light}$, effectively corresponds to exchanging $\theta_{13}$ with $\theta_{14}$. In Fig.~\ref{fig:chi2,dc3+1,2} the best-fit value $\Delta m_{41}^2=0.027~{\rm eV}^2$, shows up as a minimum in $\chi^2_{\rm DC}$. The $\bar{\nu}_e$ survival probability for the best-fit values in Eq.~(\ref{3p1bf}), is shown in Fig.~\ref{fig:prob3}. To understand qualitatively the choice of best-fit values, in Fig.~\ref{fig:his,dc3+1} we plot the energy distribution of events at Double Chooz detector. In this figure the red and blue dashed curves correspond respectively to distribution of events for the best-fit values in $3\nu$ and $(3+1)_{\rm light}$ models. The improvement of the fit to data for $(3+1)_{\rm light}$ is fairly visible, specially for prompt energies $\sim (3 - 4)$~MeV (due to the higher statistics, the main contribution comes from the left panel of Fig.~\ref{fig:his,dc3+1}). Let us take a closer look at this energy range. For example, counting the energy bins from left side, in the seventh bin ($E_{\rm prompt}\sim (3.7-4.2)$~MeV) clearly the $(3+1)_{\rm light}$ model matches the data better. This energy bin corresponds to $L/E\sim (210-233)$~m/MeV. Comparing Figs.~\ref{fig:prob11} and \ref{fig:prob3} in this range of $L/E$ shows that the $\bar{\nu}_e$ survival probability decreases in $(3+1)_{\rm light}$ model with respect to $3\nu$ framework. Numerically, for the average of probability in this bin we obtain: $\langle P \rangle_{3\nu}=0.964$ and $\langle P \rangle_{3+1}=0.883$; where the ratio of these two ($\sim 1.1$) coincides with the ratio of red to blue dashed curves in Fig.~\ref{fig:his,dc3+1,1} for the 7th bin of energy. Conversely, for the fifth bin of energy in Fig.~\ref{fig:his,dc3+1,1}, we obtain $\langle P \rangle_{3\nu}=0.947$ and $\langle P \rangle_{3+1}=0.966$ which leads to an increase in the number of events in $(3+1)_{\rm light}$ model and again better fit to data. The same improvement occurs for the other neighbor energy bins which eventually leads to a better fit in $(3+1)_{\rm light}$ model with $\Delta m_{41}^2=0.027\,{\rm eV}^2$. By changing the value of $\Delta m_{41}^2$, this distortion in the number of events moves to higher or lower energies where already the prediction of $3\nu$ model matched with data points and consequently the fit deteriorates. Notice that although a significant difference exists between Figs.~\ref{fig:prob11} and \ref{fig:prob3} in the range $L/E\sim(40 -100)$~m/MeV, since $L=1050$~m in Double Chooz experiment, contribution of this range of $L/E$, corresponding to $E_{\rm prompt} \gtrsim10$~MeV, is quite negligible.     

In Figs.~\ref{fig:allowed,dc3+1,2} and \ref{fig:allowed,dc3+1,3} we show the 2-dimensional allowed regions from Double Chooz data in $(\sin^2 2\theta_{13},\Delta m_{41}^2)$ and $(\sin^2 2\theta_{14},\Delta m_{41}^2)$ planes, respectively. As can be seen from Fig.~\ref{fig:allowed,dc3+1,2} for $\Delta m_{41}^2 \sim 0.027\,{\rm eV}^2$ even vanishing $\theta_{13}$ is allowed at 95\% C.L.. For lower values of $\Delta m_{41}^2$ ($\lesssim 0.005\,{\rm eV}^2$) also $\theta_{13}=0$ is allowed at $3\sigma$ C.L. which is a consequence of $\theta_{14}-\theta_{13}$ degeneracy mentioned in section~\ref{sec:lightsterile}. On the other hand, in the same range of $\Delta m_{41}^2$ larger values of $\sin^22\theta_{14}$ are allowed (see Fig.~\ref{fig:allowed,dc3+1,3}). At lower confidence levels, the closed allowed regions appear in both Figs.~\ref{fig:allowed,dc3+1,2} and \ref{fig:allowed,dc3+1,3} for $\Delta m_{41}^2\sim 0.027\,{\rm eV}^2$ which is a result of the minimum in Fig.~\ref{fig:chi2,dc3+1,2}.

To summarize, extending the standard $3\nu$ framework by adding one light sterile neutrino state improves the fit to Double Chooz data and weakens the lower limit on $\sin^2 2\theta_{13}$ such that the zero value is allowed at less than $\sim1\sigma$ C.L.. Also, in $(3+1)_{\rm light}$ model, the allowed region and the best-fit of $\sin^2 2\theta_{13}$ shift to smaller values. From Double Chooz data, the $(3+1)_{\rm light}$ model is favored at $\sim$ 2.2$\sigma$ C.L. with respect to $3\nu$ framework. However, it should be noticed that even in the $3\nu$ framework, the signal of nonzero $\sin^2 2\theta_{13}$ from the Double Chooz experiment is rather weak ($2.9\sigma$), and as we saw in section~\ref{sec:standard3nu}, the inclusion of Daya Bay and RENO data significantly enhance the signal. In this regard, we perform a combined analysis of all data for $(3+1)_{\rm light}$ model in the next section.

\begin{figure}[t!]
\centering
\subfloat[both reactors on]{
\includegraphics[width=0.5\textwidth]{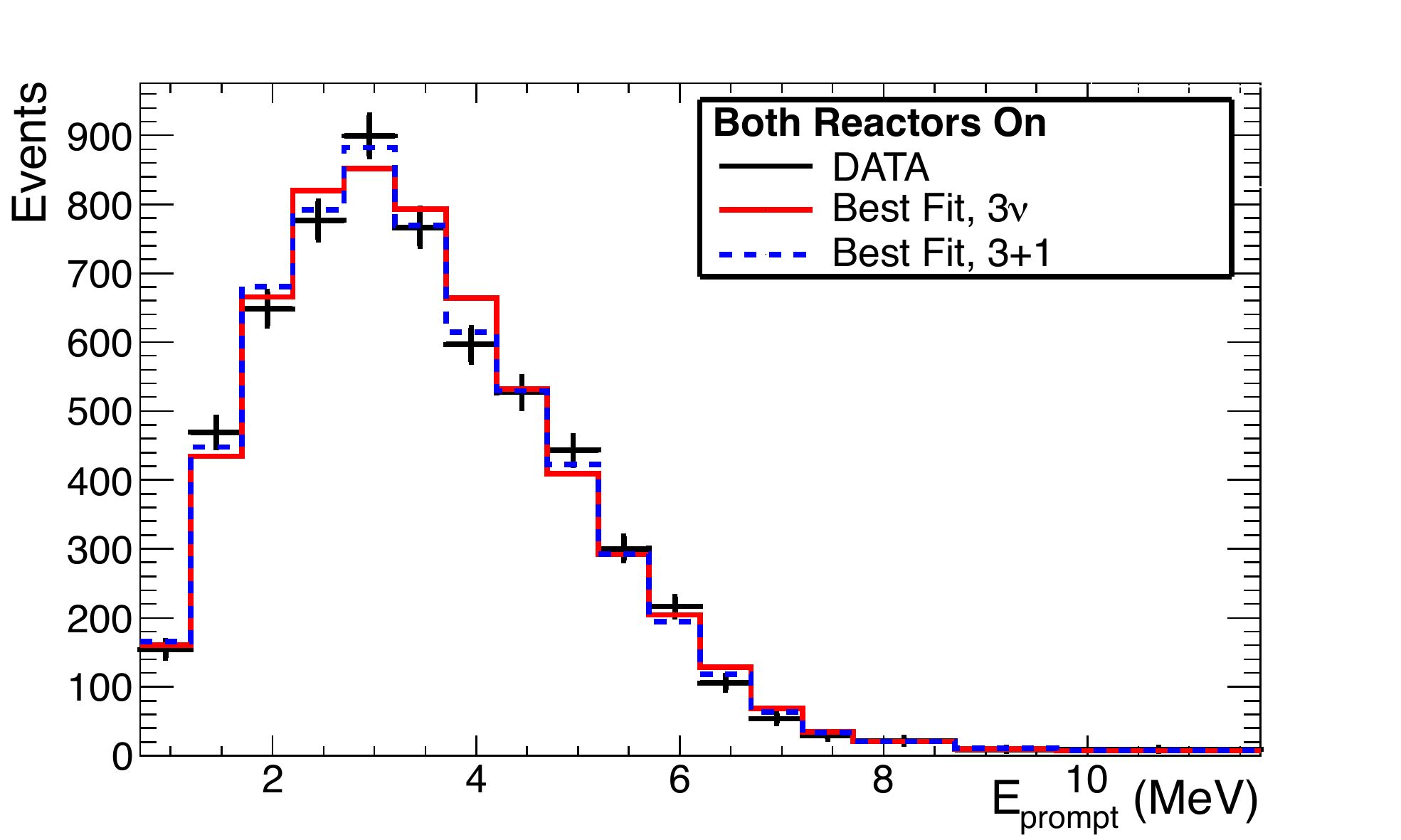}
\label{fig:his,dc3+1,1}
}
\subfloat[one reactor off]{
\includegraphics[width=0.5\textwidth]{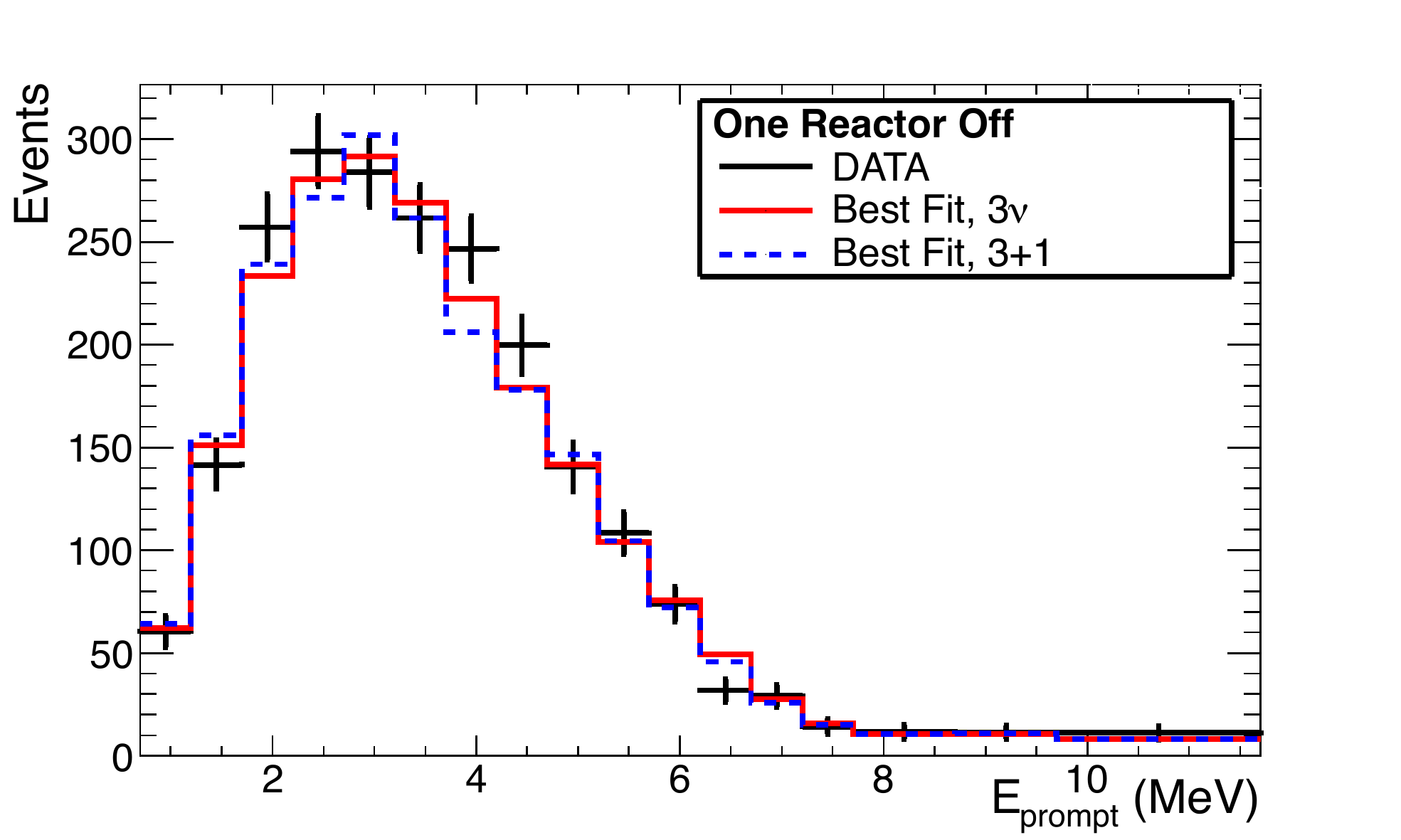}
\label{fig:his,dc3+1,2}
}
\caption{\label{fig:his,dc3+1}Prompt energy distribution of events in the Double Chooz experiment (data points), compared with the predictions of $3\nu$ (red curve) and $(3+1)_{\rm light}$ (blue dashed curve) models. Left and right panels correspond to the two data-taking periods with both reactors on and one reactor off, respectively.}
\end{figure}

\begin{figure}[t!]
\centering
\subfloat[]{
\includegraphics[width=0.5\textwidth]{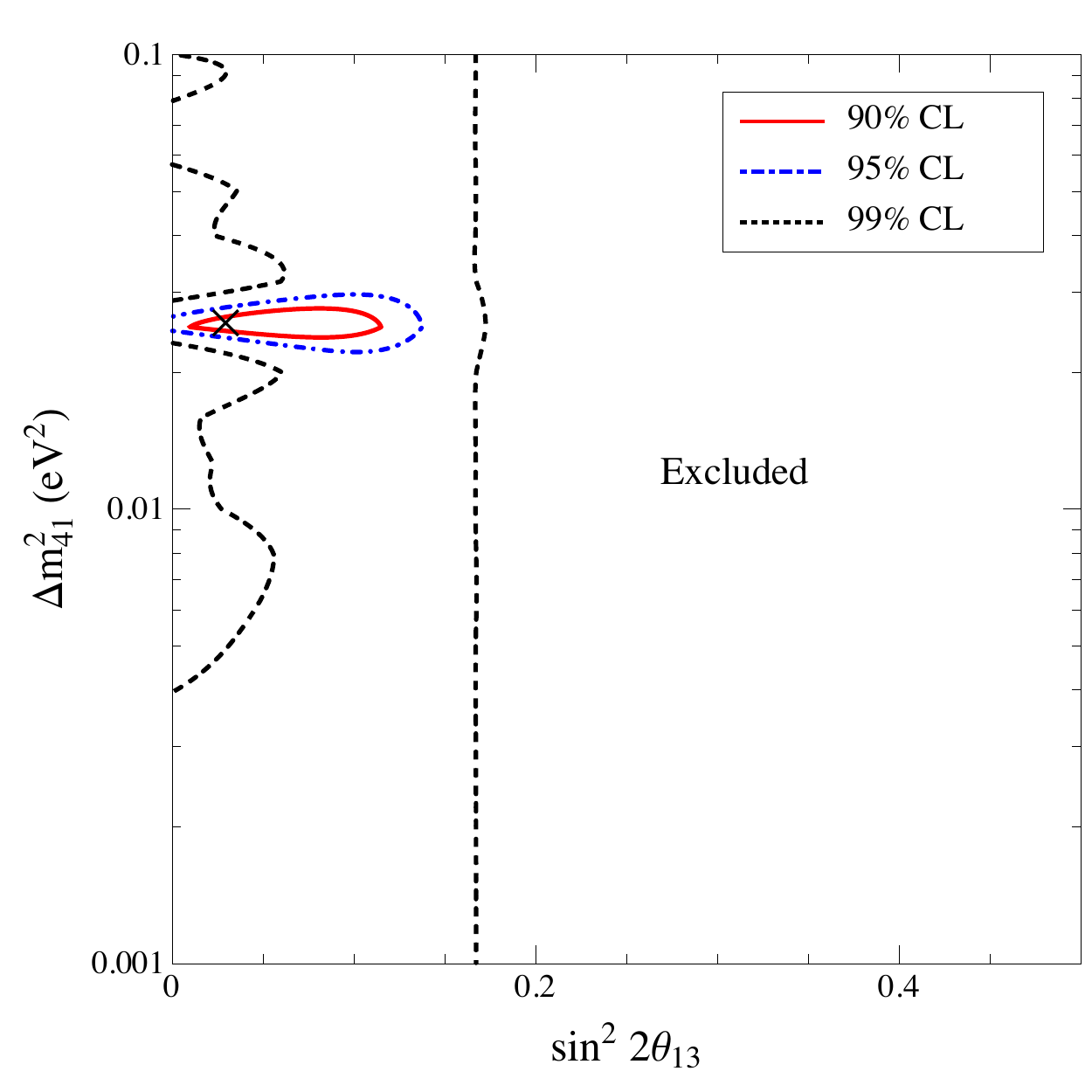}
\label{fig:allowed,dc3+1,2}
}
\subfloat[]{
\includegraphics[width=0.5\textwidth]{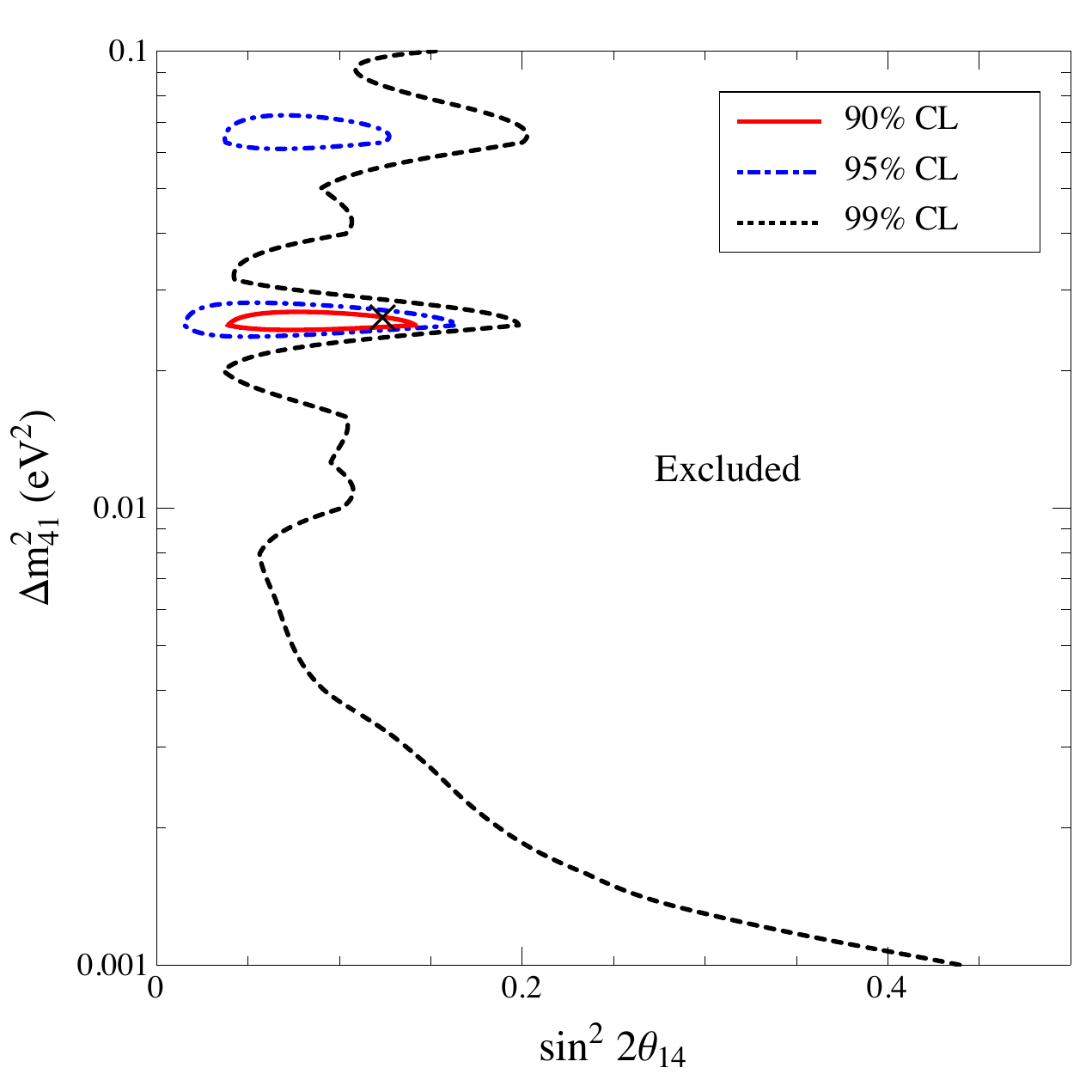}
\label{fig:allowed,dc3+1,3}
}
\caption{\label{fig:allowed,dc3+1}Allowed regions in (a) $(\sin^2 2\theta_{13},\Delta m^2_{41})$, (b) $(\sin^2 2\theta_{14},\Delta m^2_{41})$ plane from the Double Chooz data for different confidence levels. The best-fit value is shown by a cross.}
\end{figure}

\subsection{Probing $(3+1)_{\rm light}$ model with the combined data of Double Chooz, Daya Bay and RENO}
\label{combined}

In this section we probe the $(3+1)_{\rm light}$ model with the combined analysis of Double Chooz (shape and rate), Daya Bay and RENO (rate only) data. The global $\chi^2_{\rm all}$ is defined by:
\begin{equation}
\chi^2_{{\rm all}}\left(\sin^2 2\theta_{13},\sin^2 2\theta_{14},\Delta m^2_{41}\right)=\chi^2_{{\rm DC}}+\chi^2_{{\rm DB}}+\chi^2_{{\rm RENO}}~,
\end{equation}
as described in section~\ref{sec:standard3nu} in Eqs.~(\ref{chidc3}), (\ref{chidn}) and (\ref{chiRENO}). After minimizing $\chi^2_{{\rm all}}$ with respect to all the pull parameters, we find the following best-fit values
\begin{equation}
\sin^2 2\theta_{13}=0.074\quad ,\quad \sin^2 2\theta_{14}=0.059\quad ,\quad \Delta m^2_{41}=0.027~{\rm eV}^2~,
\label{3p1allexpts}
\end{equation}
with the minimum value $\chi^2_{{\rm min}}/{\rm d.o.f.}=26.7/35$. Comparing with the $3\nu$ framework $\chi^2_{\rm min}/{\rm d.o.f.}=29.7/37$, the $(3+1)_{\rm light}$ leads to improvement of the fit to global data; however, the significance of the improvement is small. 

The following comments are in order about the combined analysis: 

\begin{figure}[t!]
\centering
\subfloat[]{
\includegraphics[width=0.54\textwidth]{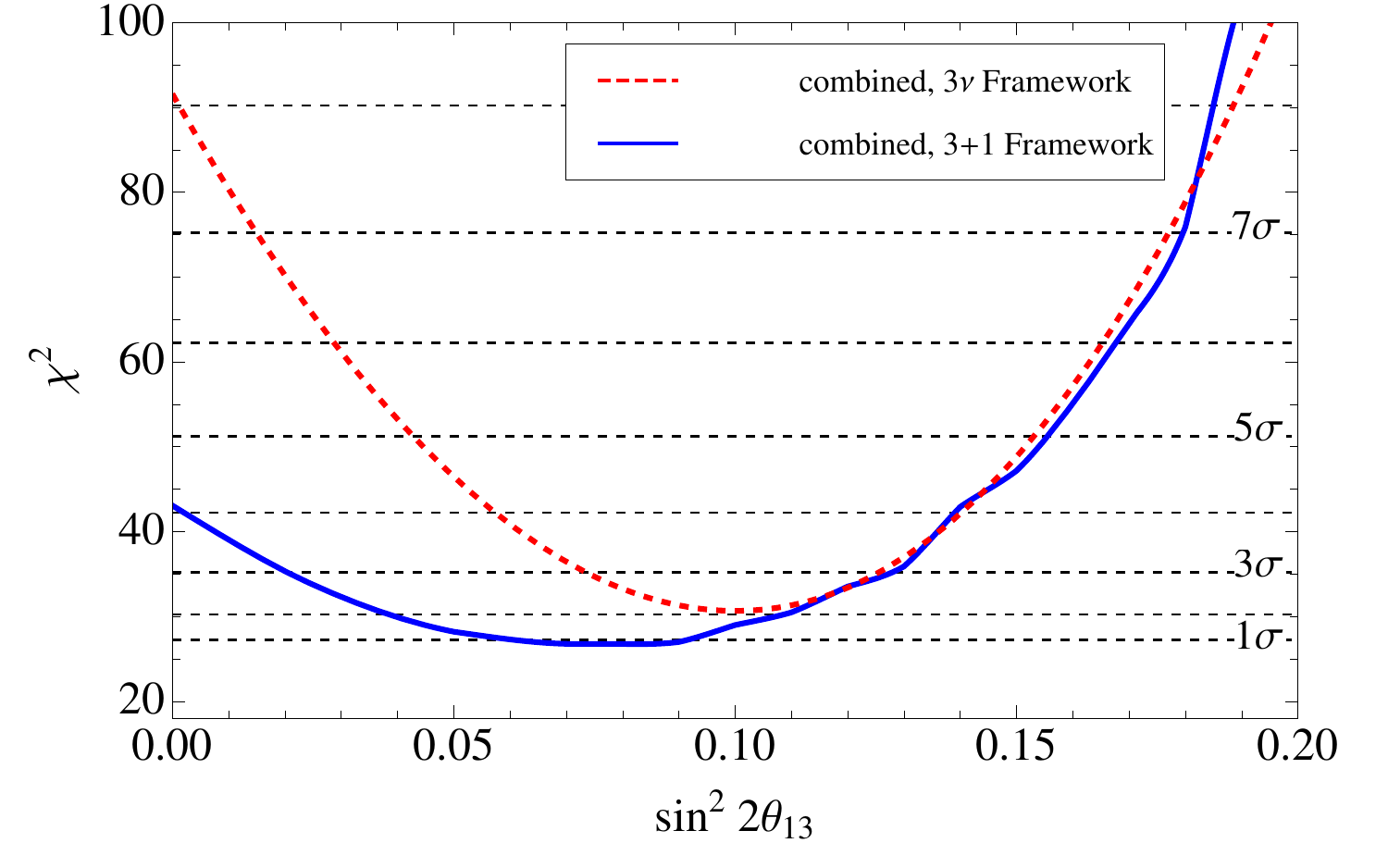}
\label{fig:chi2,combined,q13}
}
\subfloat[]{
\includegraphics[width=0.5\textwidth]{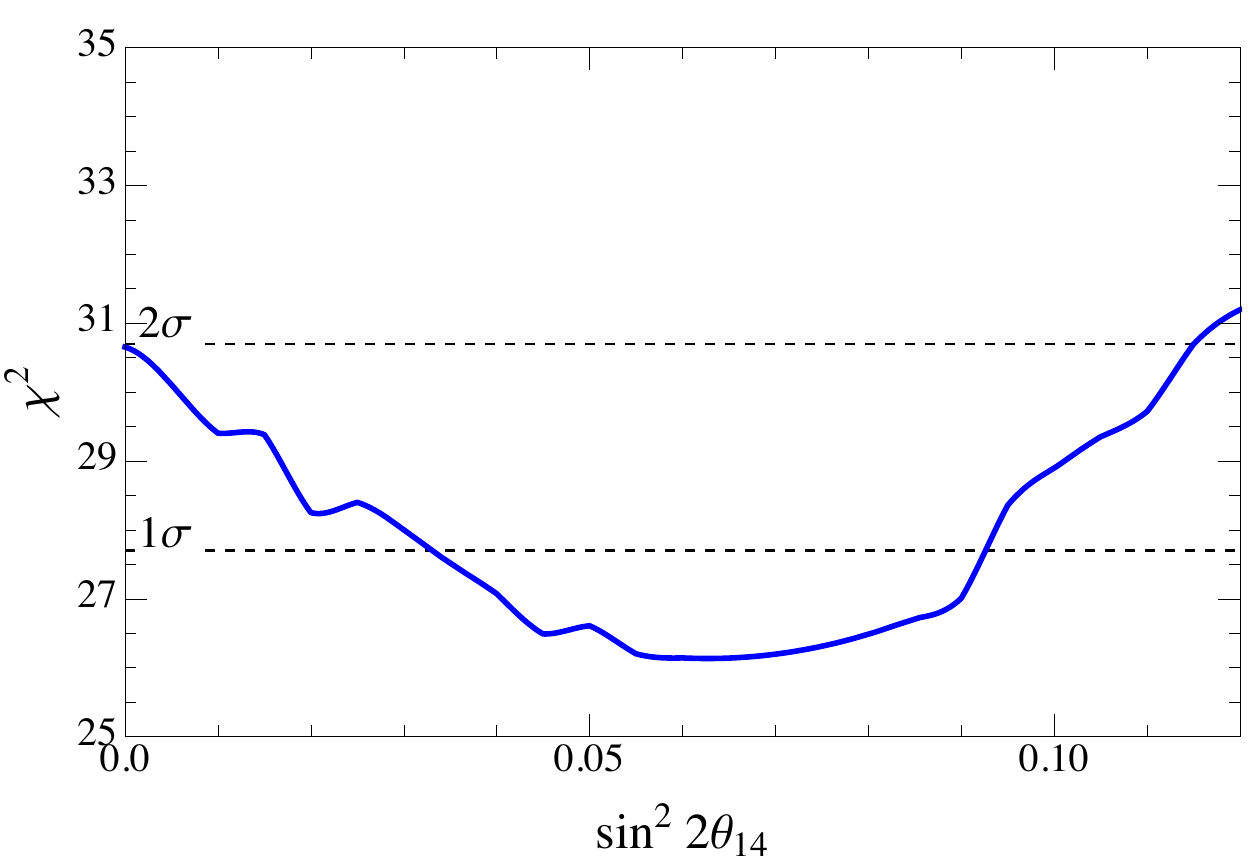}
\label{fig:chi2,combined,q14}
}
\caption{\label{fig:chi2,combined}The $\chi^2_{{\rm all}}$ versus (a) $\sin^2 2\theta_{13}$ for $(3+1)_{\rm light}$ model (blue solid curve) and $3\nu$ framework (red dashed curve) and (b) $\sin^2 2\theta_{14}$ in $(3+1)_{\rm light}$ model, for the combined data of Double Chooz, Daya Bay and RENO experiments.}
\end{figure}

\begin{itemize}

\item In Fig.~\ref{fig:chi2,combined,q13} we show $\chi^2_{\rm all}$ as a function of $\sin^2 2\theta_{13}$  for $(3+1)_{\rm light}$ model (blue solid curve) and $3\nu$ framework (red dashed curve), after marginalizing over $\Delta m_{41}^2$ and $\sin^2 2\theta_{14}$. The $1\sigma$ range of 13-mixing angle is $\sin^2 2\theta_{13}=0.074^{+0.017}_{-0.013}$. As can be seen, inclusion of Daya Bay and RENO data increases the best-fit value of $\sin^2 2\theta_{13}$, such that the zero value of $\theta_{13}$ can be excluded by $\sim 4\sigma$. The best-fit value of $\theta_{13}$ and exclusion of $\theta_{13}=0$ in the combined analysis of $(3+1)_{\rm light}$ model is comparable with the case of $3\nu$ framework, although a bit smaller (see Table~\ref{table1}).

\item Fig.~\ref{fig:chi2,combined,q14} shows $\chi^2_{\rm all}$ versus $\sin^2 2\theta_{14}$. Comparison with Fig.~\ref{fig:chi2,dc3+1,1} shows that inclusion of Daya Bay and RENO data shifts the best-fit value of $\sin^2 2\theta_{14}$ to lower values, and also the significance of nonzero $\theta_{14}$ decreases to $\sim2\sigma$. The $1\sigma$ range is $\sin^2 2\theta_{14}~=~0.059^{+0.021}_{-0.016}$.

 \begin{figure}
\begin{minipage}[b]{\textwidth}
\includegraphics[scale=0.6]{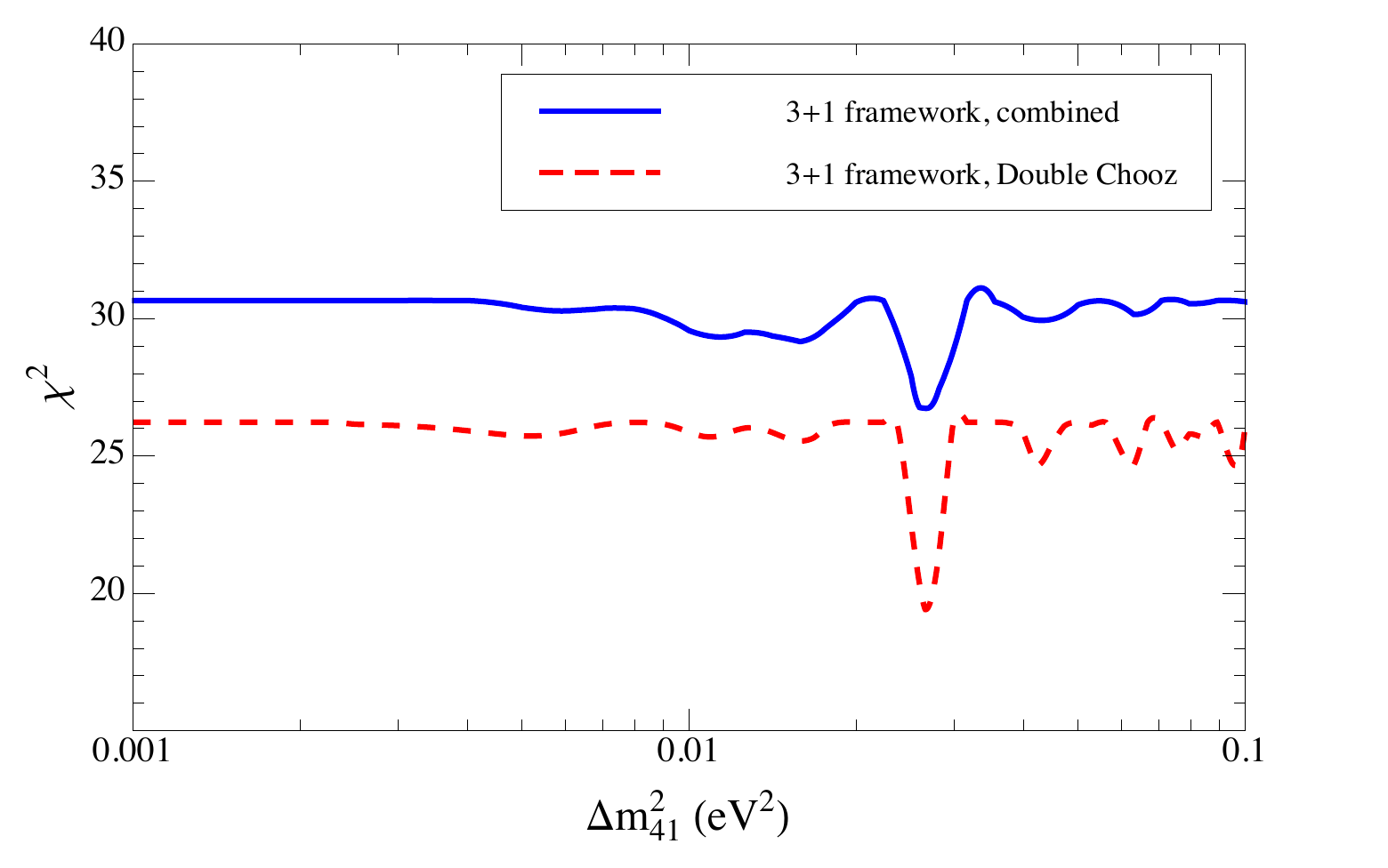}
\caption{\label{fig:combinedw}
The $\chi^2$ versus $\Delta m_{41}^2$ for the $(3+1)_{\rm light}$ model, from the combined data of Double Chooz, Daya Bay and RENO experiments (blue curve). For comparison the curve for only Double Chooz data is shown (red dashed curve).}
\end{minipage}\\
\end{figure}

\item For the mass-squared difference $\Delta m^2_{41}$, the best-fit value of Double Chooz and combined analysis is the same (see Eqs.~(\ref{3p1bf}) and (\ref{3p1allexpts}), see also Fig.~\ref{fig:combinedw}). The best-fit value $\Delta m_{41}^2=0.027\,{\rm eV}^2$ for combined analysis originates from the Double Chooz data for the same reason discussed in section~\ref{analysis3p1dc} about Fig.~\ref{fig:his,dc3+1}; namely since the position of extrema in $\bar{\nu}_e$ oscillation probability depends only on $\Delta m_{41}^2$ value. It is straightforward to check that for mixing parameters of Eq.~(\ref{3p1allexpts}) still the improvement of the fit to Double Chooz data in $(3+1)_{\rm light}$ holds to some extend in the range $E_{\rm prompt}\sim (3-4)$~MeV.  

\begin{figure}[t!]
\centering
\subfloat[]{
\includegraphics[width=0.5\textwidth]{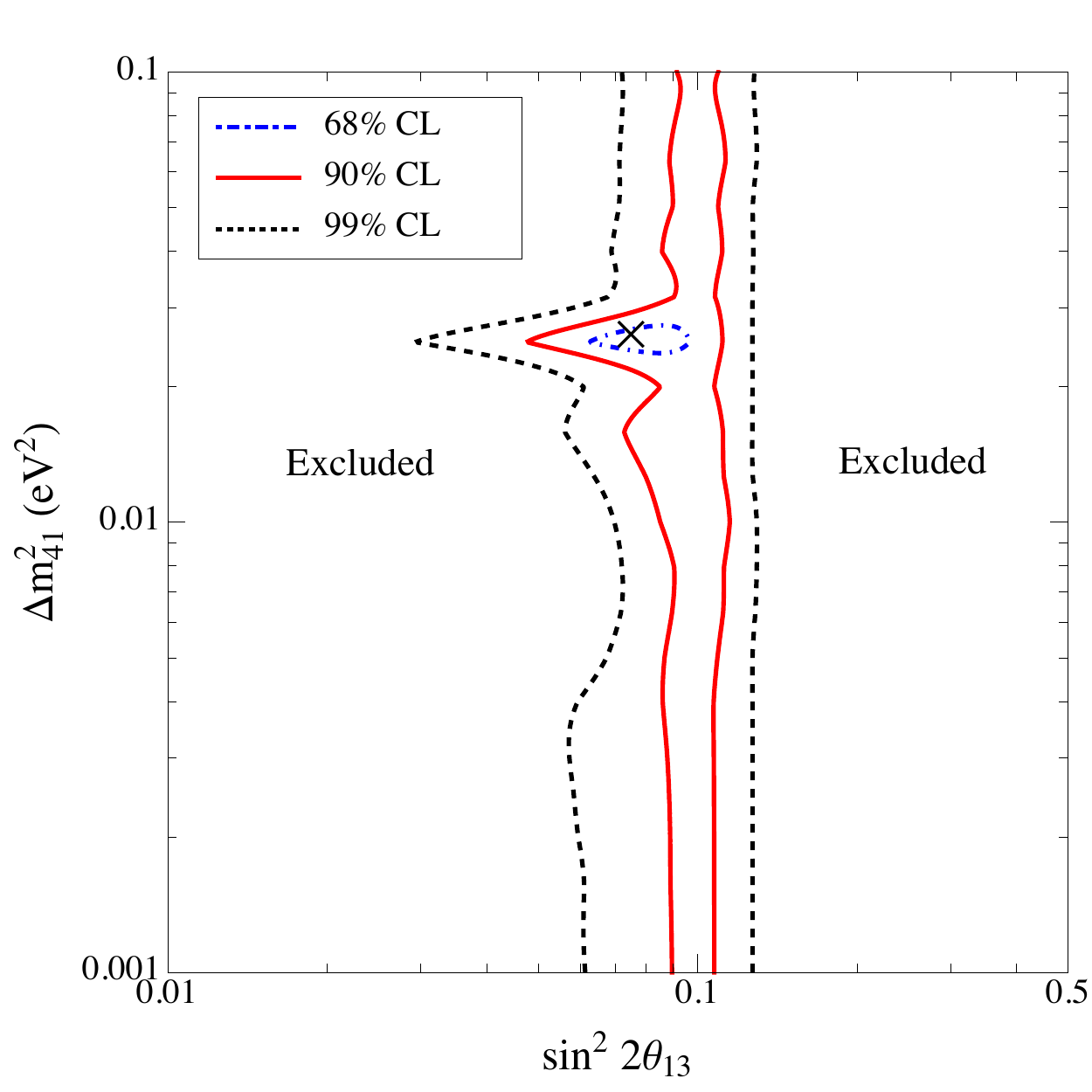}
\label{fig:allowed,combined,q13dms14}
}
\subfloat[]{
\includegraphics[width=0.5\textwidth]{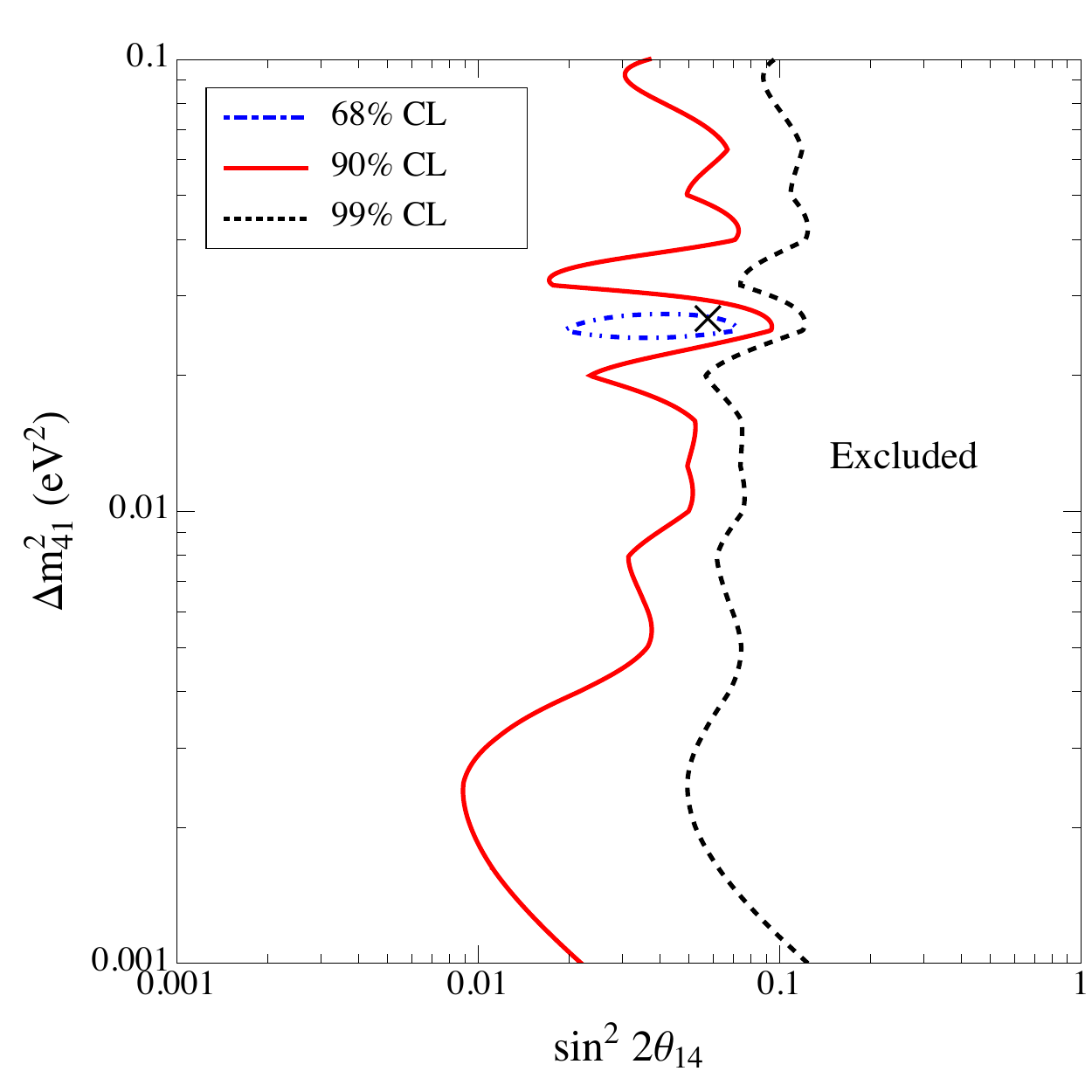}
\label{fig:allowed,combined,q14dms14}
}
\caption{\label{fig:allowed,combined,dms14}Allowed regions in (a) $(\sin^2 2\theta_{13},\Delta m^2_{41})$, (b) $(\sin^2 2\theta_{14},\Delta m^2_{41})$ plane; for the combined analysis of Double Chooz, Daya Bay and RENO data. The best-fit value is shown by a cross.}
\end{figure}

\item In Figs.~\ref{fig:allowed,combined,q13dms14} and \ref{fig:allowed,combined,q14dms14} we show the allowed region in $(\sin^2 2\theta_{13},\Delta m^2_{41})$ and $(\sin^2 2\theta_{14},\Delta m^2_{41})$ planes respectively. From Fig.~\ref{fig:allowed,combined,q13dms14} we see that the global analysis of Double Chooz, Daya Bay and RENO data in the $(3+1)_{\rm light}$ model excludes $\theta_{13}=0$ with high confidence level ($\sim4.1\sigma$). Also, the best-fit value of $\sin^22\theta_{13}$ is fairly close to the value in $3\nu$ framework. Thus, the measured value of $\sin^2 2\theta_{13}$ is robust with respect to the existence of a light sterile neutrino; and the Daya Bay and RENO data play an important role in this robustness (compare Fig.~\ref{fig:allowed,dc3+1,2} and \ref{fig:allowed,combined,q13dms14}). In both Figs.~\ref{fig:allowed,combined,q13dms14} and \ref{fig:allowed,combined,q14dms14} a closed allowed region appears at $\Delta m_{41}^2\sim 0.027\,{\rm eV}^2$ in low confidence levels which stem from the Double Chooz data. Particularly, in Fig.~\ref{fig:allowed,combined,q14dms14} the closed region indicates that $\theta_{14}=0$ can be excluded at $\sim 68\%$ C.L.; but, however, by increasing the significance the closed region transforms to upper limit and $\theta_{14}=0$ is allowed.

\item In Fig.~\ref{fig:allowed,q13q14} we show the allowed region in $(\sin^2 2\theta_{13},\sin^2 2\theta_{14})$ plane for Double Chooz only (left panel) and for the combined analysis (right panel).  In Fig.~\ref{fig:allowed,dc,q13q14} (which is for Double Chooz), there is an anti-correlation between the allowed values of $\theta_{13}$ and  $\theta_{14}$: for smaller values of $\theta_{13}$ larger values of $\theta_{14}$ are favored and vice-versa. This anti-correlation is a manifestation of $\theta_{13}-\theta_{14}$ degeneracy discussed in section~\ref{sec:lightsterile}. The break of degeneracy in low confidence levels (red and blue curves in Fig.~\ref{fig:allowed,dc,q13q14}) is due to the mismatch of Double Chooz data and $3\nu$ prediction in $E\sim(3-4)$~MeV which favors larger $\theta_{14}$. However, by including the Daya Bay and RENO data in Fig.~\ref{fig:allowed,combined,q13q14} (with the advantage of making two independent measurements in near and far detectors of each experiment) the anti-correlation and degeneracy break and end up in two islands with the characteristic that both of them favor nonzero $\theta_{13}$. For the first island: $\sin^2 2\theta_{14}<\sin^2 2\theta_{13}$, while for the second one: $\sin^2 2\theta_{14}>\sin^2 2\theta_{13}$; where the latter originates from Double Chooz contribution. In both panels of Fig.~\ref{fig:allowed,q13q14} the green dashed curve represents the limit from the solar and KamLAND data~\cite{Palazzo:2012yf}.

\begin{figure}[t!]
\centering
\subfloat[]{
\includegraphics[width=0.5\textwidth]{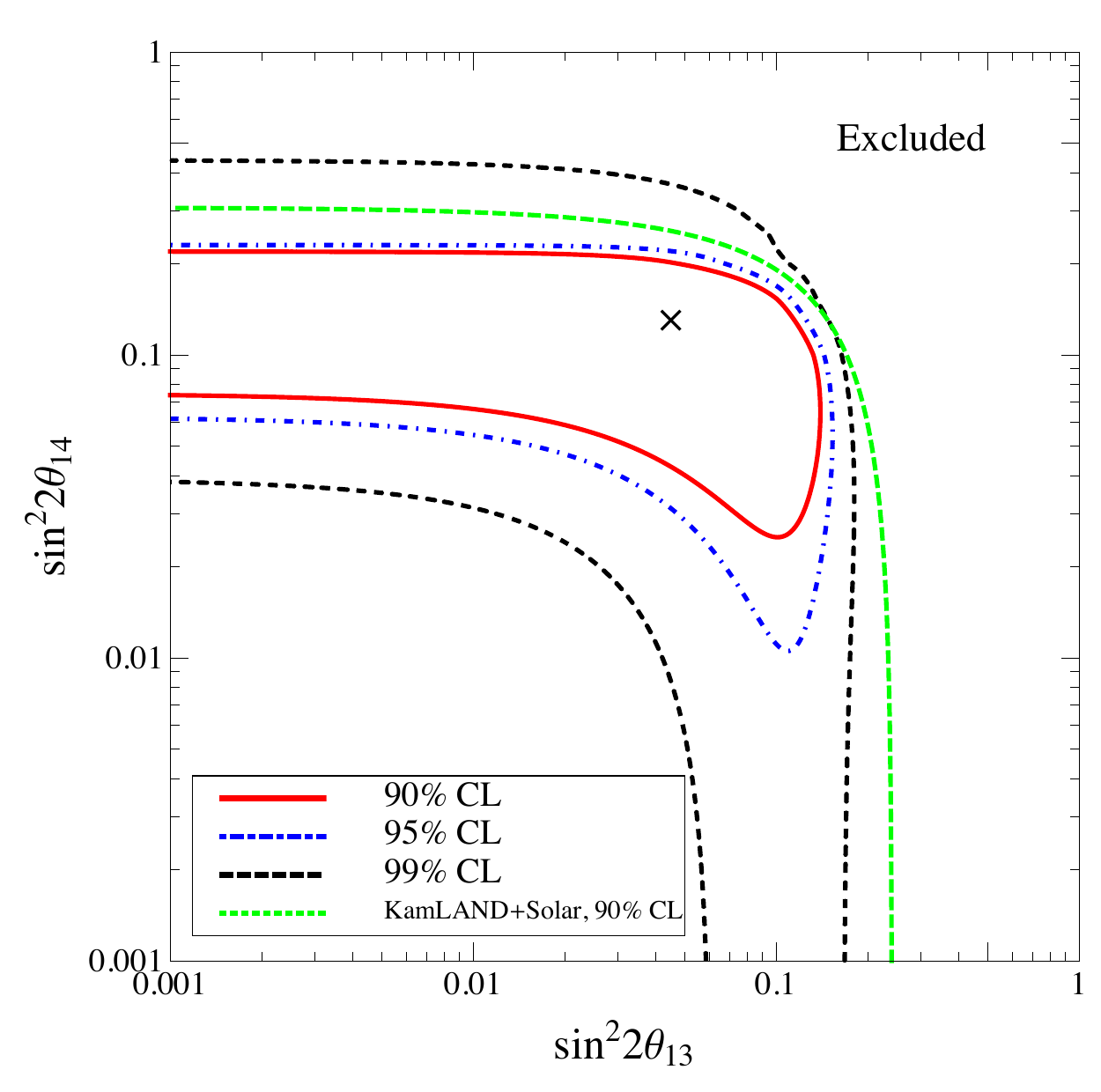}
\label{fig:allowed,dc,q13q14}
}
\subfloat[]{
\includegraphics[width=0.5\textwidth]{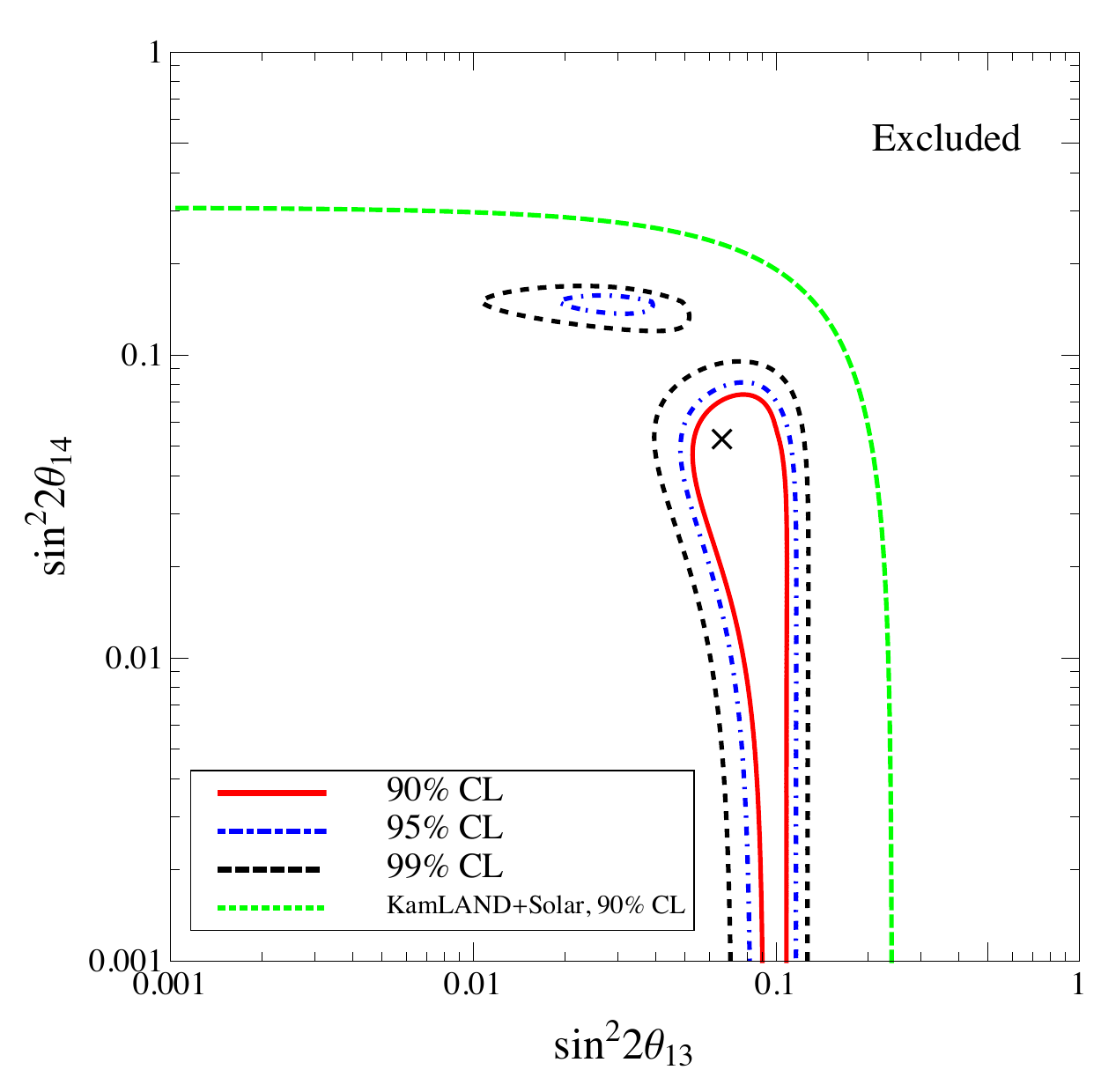}
\label{fig:allowed,combined,q13q14}
}
\caption{\label{fig:allowed,q13q14}Allowed regions in $(\sin^2 2\theta_{13},\sin^2 2\theta_{14})$ plane from Double Chooz data (left panel) and combined data of Double Chooz, Daya Bay an RENO experiments (right panel). The best-fit value is shown by a cross. The green dashed curve shows the upper limit at $90\%~$C.L. from the KamLAND and Solar data \cite{Palazzo:2012yf}.}
\end{figure}

\item Finally we would mention that our results are robust with respect to the uncertainty in the value of $\Delta m^2_{31}$, since varying this parameters will change just the positions of the extrema in Fig.~\ref{fig:prob11} while the improvement to the data requires additional extrema as in Fig.~\ref{fig:prob4}. As a cross check, we have tested the stability of our results by varying $\Delta m^2_{31}$ within their $1\sigma$ uncertainty ranges from the MINOS experiment~\cite{Adamson:2011ig}. The negligible change in our results justifies the initial assumption of fixed $\Delta m_{31}^2$.

\end{itemize}

\section{Conclusions}\label{sec:con}

Searches for the sterile neutrinos and investigating its impact on experimental results obtained, or planned to be obtained, is one of the cutting edge questions in neutrino physics. Although the initial motivation was interpretation of LSND anomaly by sterile neutrinos with mass ~$\sim\mathcal{O}(1)$~eV, further anomalies such as cosmological hints and low energy solar data stimulated the existence of lighter sterile neutrinos. In this paper we investigated the impact of existence of a light sterile neutrino, $(3+1)_{\rm light}$ model, on the medium baseline reactor experiments: Double Chooz, Daya Bay and RENO. The baseline and energy of these experiments provide the opportunity to probe active-sterile oscillation with $\Delta m_{41}^2\sim (10^{-3}-10^{-1})\,{\rm eV}^2$.  

Among these three experiments, the Double Chooz consists of one detector and present both rate and shape (in energy) information of observed events; while the Daya Bay and RENO experiments, each equipped by near and far detectors, provide the deficit in the total number of events in far detector(s) with respect to near detector(s). The energy distribution of events in Double Chooz is in good agreement with $3\nu$ framework prediction except in the range $E_{\rm prompt}\sim(3-4)$~MeV. This discrepancy leads to a better fit in $(3+1)_{\rm light}$ such that we obtained the best-fit values $\sin^2 2\theta_{13}=0.036$, $\sin^2 2\theta_{14}=0.129$ and $\Delta m^2_{41}=0.027~{\rm eV}^2$ for mixing parameters. With the Double Chooz data the $3\nu$ framework can be excluded at $\sim 2.2\sigma$ C.L.. Also, the best-fit value of $\theta_{13}$ angle is significantly different than the reported value in $3\nu$ framework and $\theta_{13}=0$ is allowed in less than $1\sigma$ C.L.. 

Inclusion of the rate information from Daya Bay and RENO experiments alters the conclusion such that with the combined data of Double Chooz, Daya Bay and RENO we obtain the best-fit values
\begin{equation}
\sin^2 2\theta_{13}=0.074 \quad , \quad \sin^2 2\theta_{14}=0.059 \quad ,\quad \Delta m^2_{41}=0.027~{\rm eV}^2~. 
\end{equation}
With the combined data the $(3+1)_{\rm light}$ model is favored at $\sim1.2\sigma$ C.L.. The value of $\theta_{13}$ angle is close to the reported value in $3\nu$ framework and so robustness of $\theta_{13}$ determination can be claimed. The persisting $\theta_{13}-\theta_{14}$ degeneracy in $(3+1)_{\rm light}$, which exists in the limit $\Delta m_{41}^2\to\Delta m_{31}^2$, can be lifted mainly from the data of Daya Bay and RENO. Despite the preference for $(3+1)_{\rm light}$ model, a large part of the parameter space of this model is excluded in our analysis, better than the other constraints by a factor of 2.

The planned near future data from these experiments can significantly exclude/strengthen the favored nonzero active-sterile mixing parameters found in this paper. Especially, the energy spectrum of data in Daya Bay and RENO experiments can decisively rule out/confirm it. Also, installation of near detector in the Double Chooz experiment can provide valuable information about the observed anomaly in $E_{\rm prompt}\sim(3-4)$~MeV, wether supporting it or contradicting it.

{\it Note added 1:} After completion of this work, the energy spectrum of Daya Bay data have been reported~\cite{dayabayspectrum}. Interestingly, an anomalous behavior, similar to the Double Chooz data, in the energy range $(3-4)$~MeV can be identified. However, a detailed analysis taking into account both Daya Bay and Double Chooz data is worthwhile.

{\it Note added 2:} Before submitting this paper, the article~\cite{Palazzo:2013bsa} appeared on arXiv with results close to the results in this paper. However, in~\cite{Palazzo:2013bsa} just the rate analysis of the Double Chooz data is considered and consequently the $(3-4)$~MeV anomaly mentioned in this paper does not show up in~\cite{Palazzo:2013bsa}.

\begin{acknowledgments}
O.~L.~G.~P. thanks the ICTP and the financial support from the funding grant 2012/16389-1, S\~ao Paulo Research Foundation (FAPESP). 
A. E. thanks the financial support from the funding grant 2009/17924-5, S\~ao Paulo Research Foundation (FAPESP). E.K. thanks support from CNPq (GRANT 311102/2011-5 ) and FAPESP (GRANT 2010/07359-6). We would like to thank Antonio Palazzo for providing us the information about KamLAND+Solar limit.
\end{acknowledgments}


\end{document}